\documentclass[aps,pra,twocolumn,showpacs]{revtex4}
\usepackage{epsf}
\usepackage{amsmath}
\usepackage{bm}

\begin{document}

\title{Field quantization for open optical cavities}
\date{\today}

\author{
Carlos Viviescas and
Gregor Hackenbroich}
\affiliation{
Universit\"at Essen, Fachbereich Physik,
 45117 Essen, Germany}

\begin{abstract}
  We study the quantum properties of the electromagnetic field in
  optical cavities coupled to an arbitrary number of escape
  channels. We consider both inhomogeneous dielectric resonators with
  a scalar dielectric constant $\epsilon({\bf r})$ and cavities
  defined by mirrors of arbitrary shape. Using the Feshbach projector
  technique we quantize the field in terms of a set of resonator and
  bath modes. We rigorously show that the field Hamiltonian reduces to
  the system--and--bath Hamiltonian of quantum optics. The field
  dynamics is investigated using the input--output theory of Gardiner
  and Collet. In the case of strong coupling to the external radiation
  field we find spectrally overlapping resonator modes. The mode
  dynamics is coupled due to the damping and noise inflicted by the
  external field. For wave chaotic resonators the mode dynamics is
  determined by a non--Hermitean random matrix. Upon including an
  amplifying medium, our dynamics of open-resonator modes may serve as
  a starting point for a quantum theory of random lasing.
\end{abstract}


\maketitle

\section{Introduction}
\label{sec:intro}

In the past few years advances in microstructuring techniques have
made it possible to manufacture novel mirrorless cavities known as
random lasers. Feedback in these lasers is provided by the random
scattering of light in a medium with spatially fluctuating refractive
index. Random lasing has been demonstrated in semiconducting clusters
and films \cite{Cao99} as well as in solutions of $\textrm{TiO}_2$
nanoparticles \cite{Cao00,Lin02} and in polymer systems
\cite{Fro98}. Random lasers typically have dimensions of several $\mu
m$ and may prove useful in microtechnology applications \cite{Wie00}.
Fundamentally, interest in these systems derives from the complexity
of the interplay of amplification, disorder and random scattering of
light.

The theoretical investigation of disordered media has a long history
\cite{She90,She95}, but most studies are restricted to {\em passive}
systems in the regime of {\em classical optics}.  Far reaching
similarities have been revealed between the propagation of light
through (passive) random dielectrics and the transport of electrons
through disordered solids.  Well-known manifestations of this
similarity are the coherent backscattering of light \cite{Alb85}
versus the weak--localization for electrons \cite{Imr97} and the
strong localization of microwaves \cite{Gen91} or light \cite{Wie97}
versus the Anderson localization of electrons \cite{And58}.
Conceptually, this similarity may be understood from the nonlinear
supersymmetric $\sigma$ model which proved to be the underlying field
theory both for disordered solids \cite{Efe97} and for classical
optical waves in random dielectrics \cite{Ela98}.

In this paper we address the {\em quantum} properties of the
electromagnetic field in the presence of a random medium. A quantum
treatment of light is required when one wants to compute the
spectrum, the linewidth or the photon statistics of the output
radiation. We study both dielectric media with a spatially
nonuniform dielectric constant $\epsilon({\bf r})$ and optical
cavities defined by mirrors of arbitrary shape. These systems may be
coupled to an arbitrary number of escape channels. A special case
are cavities that, e.g.\ due to partly reflecting mirrors, are only
weakly coupled to the external radiation field.  Such cavities are
used in standard lasers, and have been extensively studied in laser
theory. The standard quantum theory of the laser \cite{Hak84} starts
from the expansion of the electromagnetic field in terms of closed
resonator modes. Escape from the resonator gives rise to a small
perturbation of the closed resonator dynamics. To leading order in a
perturbation theory in terms of the coupling to the outside world,
the resonator modes may well be approximated by modes of an entirely
closed system.  This perturbation theory breaks down in cavities
with a weak confinement of light. For such systems, the field
quantization of standard resonator theory must be replaced by a
generalization suitable for open resonators. We present and discuss
such a generalized quantization technique in this paper, and
demonstrate that it is well suited for analytical investigations of
random media.

Our solution of the quantization problem is most relevant for ``open''
cavities like random lasers, but is significant for a more fundamental
reason. Our technique allows for a rigorous treatment of a fundamental
problem of quantum optics, concerning the damping of radiation due to
leakage out of a cavity.  Conventionally, this problem has been
described by a system--and--bath Hamiltonian \cite{Sen59,Gar00}: The
system is modeled by a discrete set of independent quantized harmonic
oscillators associated with the normal modes of a completely isolated
cavity. A different, continuous set of oscillators represents the bath
while a coupling between the system and bath oscillators gives rise to
damping. In spite of its intuitive appeal and its success in
applications, this model of damping has been criticized. The
formulations \cite{Sen59,Gar00} lack a microscopic justification as
the coupling constants only enter as phenomenological parameters.
Moreover, it has been argued \cite{Bar88,Dut00a} that the approach is
restricted to good resonators with damping rates much smaller than the
mode frequency spacing. We show below that such pessimism is
inappropriate: We rigorously derive the system--and--bath Hamiltonian
starting from the Maxwell equations. We employ a technique from
nuclear and condensed matter physics known as the Feshbach projector
formalism. Expressions are obtained for the coupling amplitudes. We
find that the field Hamiltonian generally includes both resonant and
non--resonant terms. When both are kept, the resulting dynamics
correctly describes damping and even overdamping (when the damping
rate becomes of the order of the frequency).

We note that there are alternative approaches to the field
quantization in open resonators. These are either based on mode
expansions of the electromagnetic field or they altogether abandon the
notion of cavity modes and directly quantize the wave equation using
Green functions. Approaches based on mode expansions include the
modes--of--the--universe or true--mode approach (the field is expanded
in terms of eigenstates of the total system comprising the resonator
and the bath) \cite{Lan73,Gla91,Kno87}, expansions in terms of
so--called quasi--modes \cite{Dal99} or in terms of non--orthogonal
(or Fox--Li) modes \cite{You98,Lam99,Lam02,Dut00b,Bro01} (a
characterization of these approaches can be found in Ref.\
\cite{Dut00b}). Alternative approaches based on Green functions were
developed by Grunder and Welsch \cite{Gru96} and by Loudon and
co--workers \cite{Art97}. So far, to our knowledge, none of the first
approaches has been applied to random media. In contrast, the second
(Green function) approach has been successfully used by Beenakker and
coworkers \cite{Bee98,Mis00} to establish a general relation between
the emission of {\em linear} optical media and the underlying
scattering matrix.  Combining this relation with the statistical
properties of the scattering matrix in disordered media, Beenakker
derived the full photocount distribution of the radiation emitted from
linear random media.  Unfortunately, this random scattering approach
suffers from a fundamental limitation: It is restricted to {\em
linear} media and therefore cannot describe lasers above the lasing
threshold. No such limitations hold for the approach presented in this
paper as this approach is based on an Hamiltonian: atom--field
interactions can be included in a standard way.

The outline of the paper is a follows. In Sec.\ \ref{sec:fieldquant}
we describe the field quantization. We start from a global expansion
of the field in terms of the eigenmodes of the Maxwell equations.  The
system--and--bath Hamiltonian is then derived using Feshbach's
projector technique. Our derivation is valid for spatially
inhomogeneous dielectrics and fields of arbitrary polarization. This
is a substantial generalization of our previous work \cite{Hac01}
which was restricted to scalar fields and homogeneous dielectric
media. In Sec.~\ref{sec:linear} we investigate the field
dynamics. Using the input--output theory of Gardiner we calculate the
output field in the presence of an arbitrary number of escape
channels. We derive the equations--of--motion of the internal cavity
modes, and show that the dynamics of these modes is coupled by damping
and noise due to the external radiation field. As a final illustration
of our technique we compute in Sec.~\ref{sec:spont} the decay rate of
a single two--level atom in a cavity of arbitrary shape. We conclude
by discussing possible further applications of our technique, most
notably the application to random lasers.

\section{Field quantization}
\label{sec:fieldquant}

\subsection{Normal modes}
\label{sec:normal}

We consider a three dimensional linear dielectric me\-dium
characterized by a scalar dielectric constant $\epsilon({\bf r})$
that depends explicitly on position. We assume that the dielectric
constant is real and frequency independent. The dielectric is
surrounded by free space. Cavities defined by (ideal) mirrors are a
special case with $\epsilon({\bf r}) \equiv 1$ and appropriate
boundary conditions on the mirrors. The electromagnetic field for
the total system, comprising the resonator and the external
radiation field, may be quantized using the exact eigenmodes of
Maxwell's equations.  This so--called modes--of--the--universe
approach \cite{Lan73,Gla91} serves as a starting point for the
derivation of the system--and--bath Hamiltonian in
Sec.~\ref{sec:hamil}; we therefore summarize the main steps of this
approach below.

It is convenient to formulate the quantization procedure in terms of
the vector potential ${\bf A}$ and the scalar potential $\phi$. We
work in the Coulomb gauge which, in the absence of sources,
corresponds to the choice $\phi = 0$ and the generalized
transversality condition $\nabla \cdot [\epsilon({\bf r}) {\bf A}] =
0$.  The magnetic and electric field then follow from the potentials
via the familiar relations
\begin{equation}
\label{eq:ebfield}
{\bf E} = - \frac{1}{c} \frac{\partial {\bf A}}{\partial t},
\qquad {\bf B} = \nabla \times {\bf A}.
\end{equation}
The electromagnetic Hamiltonian of the problem is given by
\begin{eqnarray}
\label{eq:huniv}
H = \frac{1}{2} \int d {\bf r} \left[ \frac{c^2 {\bm \Pi}({\bf
      r},t)^2}{\epsilon({\bf r})} + (\nabla \times {\bf A}({\bf 
     r},t))^2
\right],
\end{eqnarray}
where ${\bm \Pi}({\bf r})= \epsilon({\bf r}) \dot{\bf A}({\bf
  r})/c^2$ is the canonical momentum field.  The quantization of the
fields may be achieved by imposing a suitable commutation relation
between ${\bf A}({\bf r},t)$ and ${\bm \Pi}({\bf r},t)$. An
alternative but equivalent procedure is to expand the fields in a
complete set of mode functions and to impose canonical commutation
relations for the expansion coefficients. We follow the second
procedure here, and expand the vector potential in terms of the
exact eigenmodes ${\bf f}_m (\omega,{\bf r})$, defined as solutions
of the wave equation
\begin{eqnarray}
\label{eq:evequniv}
\nabla \times \left[ \nabla \times {\bf f}_m(\omega,{\bf r})
  \right] - \frac{\epsilon({\bf r})\omega^2}{c^2} {\bf
  f}_m(\omega,{\bf r})={\bm 0} .
\end{eqnarray}
The solutions automatically satisfy the transversality condition $
\nabla \cdot \left[\epsilon({\bf r}) {\bf f}_m(\omega, {\bf r})
\right] = 0$.  The eigenmodes are labeled by the continuous
frequency $\omega$ and a discrete index $m$ that specifies the
asymptotic boundary conditions far away from the dielectric
(including the polarization). We consider asymptotic conditions
corresponding to a scattering problem with incoming and outgoing
waves.  Then ${\bf f}_m(\omega,{\bf r})$ represents a solution with
an incoming wave in channel $m$ and only outgoing waves in all other
scattering channels. The definition of the channels depends on the
problem at hand: For a dielectric coupled to free space, one may
expand the asymptotic solutions in terms of angular momentum states.
Then $m$ corresponds to an angular momentum quantum number. On the
other hand, for a dielectric connected to external waveguides, $m$
may represent a transverse mode index.  It is convenient to combine
the solutions associated with the different channels to an
$M$--component vector ${\bf f}(\omega,{\bf r})$ where $M$ is
the total number of open channels at frequency $\omega$. The field
expansions then take the form
\begin{subequations}
\label{eq:qpexp}
\begin{eqnarray}
{\bf A}({\bf r},t) & = & c \int \! d\omega \;  q (\omega,t)
{\bf f} (\omega,{\bf r})  , \\ 
{\bm \Pi}({\bf r},t) & = & \frac{1}{c} \int \! d\omega \; 
{\bf f}^\dagger (\omega,{\bf r}) p (\omega,t) ,
\end{eqnarray}
\end{subequations}
where the expansion coefficients $q(\omega)$ and $p(\omega)$ form
$M$--component row and column vectors, respectively. The components
$q_m(\omega, t)$, $p_m(\omega, t)$ are time--dependent variables.

The eigenmodes fulfill an orthonormality relation. This follows from
the observation that the substitution
\begin{equation}
\label{eq:fguniv}
{\bf f} (\omega,{\bf r}) = \frac{1}{\sqrt{\epsilon({\bf r})}} \bm{\phi} 
(\omega,{\bf r})
\end{equation}
transforms Eq.~(\ref{eq:evequniv}) into the eigenvalue problem
\begin{eqnarray}
\label{eq:hevequniv}
L \bm{\phi}  (\omega, {\bf r}) \equiv
\frac{1} {\sqrt{\epsilon({\bf r})}} \nabla \times 
\left[ \nabla \times \frac{ \bm{\phi} (\omega, {\bf r})}
{\sqrt{\epsilon({\bf r})}} \right] =
\frac{\omega^2} {c^2} \bm{\phi} (\omega, {\bf r}) ,
\end{eqnarray}
for the Hermitean differential operator $L$. Choosing an orthonormal
set of basis functions $\bm{\phi}_m(\omega, {\bf r})$, it follows
that the associated mode functions ${\bf f}_m (\omega)$ satisfy the
orthonormality condition
\begin{equation}
\label{eq:forth}
\int \! d{\bf r} \; \epsilon({\bf r}) {\bf f}^*_m(\omega,{\bf r}) 
\cdot {\bf f}_{m^\prime}(\omega^\prime,{\bf r}) = \delta_{mm^\prime} \, 
\delta(\omega - \omega^\prime).
\end{equation}
The functions ${\bm \phi}_m(\omega)$ form a complete set in the
subspace of $L^2$ functions defined by the transversality
condition
\begin{equation}
\label{eq:gauge}
\nabla \cdot \left[\sqrt{\epsilon({\bf r})} {\bm \phi}_m 
({\bf r}) \right] = 0.
\end{equation} 
The associated mode functions ${\bf f}_m(\omega)$ form a complete
set in the space of transverse functions \cite{Gla91}.

Inasmuch as the fields are real, the vector potential and its
canonical momentum fulfill the relations ${\bf A}={\bf A}^{\dagger}$
and ${\bm \Pi}={\bm \Pi}^{\dagger}$. Together with
Eq.~(\ref{eq:qpexp}), this implies
\begin{subequations}
\label{eq:qpreal}
\begin{eqnarray}
q_m(\omega) &=& \sum_{m^\prime} \int \! d\omega^\prime \; 
\mathcal{M}_{mm^\prime}^{\dagger}(\omega,\omega^\prime)
q_{m^\prime}^{\dagger} (\omega^\prime), \\ 
p_m^{\dagger}(\omega) &=& \sum_{m^\prime} \int \! d\omega^\prime \;
\mathcal{M}_{mm^\prime}^{\dagger}(\omega,\omega^\prime)
p_{m^\prime}(\omega^\prime),
\end{eqnarray}
\end{subequations}
where $\mathcal{M}$ has the matrix elements
\begin{eqnarray}
\label{eq:uuniv}
\mathcal{M}_{m m^\prime}(\omega,\omega^\prime) = \int \! d{\bf r} 
\; \epsilon ({\bf r}) {\bf f}_{m}(\omega,{\bf r}) \cdot 
{\bf f}_{m^\prime} (\omega^\prime,{\bf r}).
\end{eqnarray}
We note that $\mathcal{M}$ is unitary and symmetric \cite{foot1}.
Moreover, $\mathcal{M}$ only couples degenerate modes, as modes
with different frequencies are orthogonal, $\mathcal{M}(\omega,
\omega^\prime) \sim {\delta(\omega-\omega^\prime)}$.

Substituting the field expansions (\ref{eq:qpexp}) into the
Hamiltonian (\ref{eq:huniv}), and using Eq.~(\ref{eq:qpreal}) and
the properties of $\mathcal{M}$, one obtains the Hamiltonian in
terms of the variables $q$ and $p$,
\begin{eqnarray}
\label{eq:hqp}
H = \frac{1}{2} \sum_m \int \! d\omega \; \left[ p_m^\dagger(\omega) 
p_m(\omega) + \omega^2 q_m^{\dagger}(\omega) q_m(\omega) 
\right]. \quad
\end{eqnarray}
Quantization is now achieved by promoting the variables $q(\omega)$
and $p(\omega)$ to operators.  The Heisenberg equations of motion
for $q(\omega)$ and $p(\omega)$ lead to Maxwell's equations,
provided we impose the equal time commutation relations
\begin{eqnarray}
\label{eq:qpcomm}
\ [q_m(\omega),q_n(\omega^\prime)] &=&
[q_m(\omega),q_n^{\dagger}(\omega^\prime)] = 0, \nonumber \\ 
\ [p_m(\omega),p_n(\omega^\prime)] &=&
[p_m(\omega),p_n^{\dagger}(\omega^\prime)] =0, \\
\ [q_m(\omega),p_n(\omega^\prime)] &=& i\hbar \, \delta_{mn} 
\, \delta(\omega - \omega^\prime) \nonumber.
\end{eqnarray}
Further use of Eq.~(\ref{eq:qpreal}) gives the remaining commutation
relation
\begin{eqnarray}
\label{eq:qpdcom}
[q_m(\omega),p^{\dagger}_n(\omega^\prime)] &=& i\hbar
\mathcal{M}^{\dagger}_{mn}(\omega,\omega^\prime).
\end{eqnarray}
Combining the field expansions (\ref{eq:qpexp}) with the
orthogonality condition (\ref{eq:forth}), one can show that these
commutation relations imply canonical commutation relations for the
vector potential and the canonical momentum field.

The last step in the quantization procedure is to express the
operators $q(\omega)$ and $p(\omega)$ in terms of creation and
annihilation operators,
\begin{subequations}
\label{eq:qpaexp}
\begin{eqnarray}
q_m(\omega) &=& \left[  \frac{\hbar}{2\omega} \right]^{\frac{1}{2}}
  \bigg[  A_m(\omega)  \nonumber \\
 &&  + \sum_{m^\prime} \int \! d\omega^\prime \;
  \mathcal{M}_{mm^\prime}^{\dagger}(\omega,\omega^\prime)
  A_{m^\prime}^{\dagger} (\omega^\prime)  \bigg], \quad \quad 
\end{eqnarray}
\begin{eqnarray}
p_m(\omega) &=& i \left[ \frac{\hbar \omega}{2}
  \right]^{\frac{1}{2}} \bigg[ A_m^{\dagger}(\omega) \nonumber \\
&&  - \sum_{m^\prime} \int \! d\omega^\prime \; 
  \mathcal{M}_{mm^\prime}(\omega,\omega^\prime) 
   A_{m^\prime}(\omega^\prime)
  \bigg]. \quad \quad
\end{eqnarray} 
\end{subequations}
The latter obey the commutation relations
\begin{eqnarray}
\ [A_m(\omega), A_{m^\prime}(\omega^\prime)] &=& 0, \nonumber \\
\ [A_m(\omega), A^{\dagger}_{m^\prime}(\omega^\prime)] &=&
\delta_{mm^\prime} \, \delta(\omega - \omega^\prime).
\end{eqnarray}
In terms of the creation and annihilation operators the Hamiltonian
takes the familiar form 
\begin{eqnarray}
\label{eq:ha}
H = \frac{1}{2} \sum_m \int \! d\omega \, \hbar\omega \left[ A_m^\dagger(\omega) 
A_m(\omega) + A_m(\omega) A_m^\dagger(\omega) \right], \quad
\end{eqnarray}
describing a set of independent harmonic oscillators. Finally,
substituting the representations (\ref{eq:qpaexp}) into
Eqs.~(\ref{eq:qpexp}), one obtains the field expansions
\begin{subequations}
\label{eq:aexp}
\begin{eqnarray}
{\bf A}\! & = & \!\!c \sum_m \int \! d\omega \; \left[
  \frac{\hbar}{2\omega}\right]^{\frac{1}{2}} \left[ A_m(\omega,t) {\bf
    f}_m(\omega,{\bf r}) + \mbox{h.c.} \right], \quad \quad \\
{\bm \Pi}\! & = &\!\! -\frac{i}{c} \sum_m \int \! \! d\omega \left[
  \frac{\hbar\omega}{2}\right]^{\frac{1}{2}} \left[ A_m(\omega,t)
  {\bf f}_m(\omega,{\bf r}) - \mbox{h.c.} \right], \quad \quad
\end{eqnarray}
\end{subequations}
In empty space, $\epsilon({\bf r}) \equiv 1$, they reduce to the
standard mode expansion of the free electromagnetic field.

\subsection{Feshbach--projection}
\label{sec:hamil}
The modes--of--the--universe approach yields a consistent
quantization scheme for the electromagnetic field, but does not
provide explicit information about the field inside the resonator.
The dynamics of this field is important in various contexts:
Long--lived resonator modes are responsible for scattering
resonances and, in the presence of an amplifying medium, may turn
into lasing modes. To introduce resonator modes and to discuss their
dynamics, we now separate the electromagnetic field into two
contributions, accounting, respectively, for the field inside and
outside the resonator. Formally, the separation of space in two
regions is achieved using the projection operators \cite{Fes62}
\begin{subequations}
\label{eq:defq}
\begin{eqnarray}
\mathcal{Q} &=& \int_{{\bf r} \in I} \! d{\bf r} |{\bf r} 
\rangle\langle{\bf r}|, \\
\mathcal{P} &=& \int_{{\bf r} \not\in I} \! d{\bf r} |{\bf r}
\rangle\langle{\bf r}|,
\end{eqnarray}
\end{subequations}
where $|{\bf r} \rangle$ denotes a standard position eigenket and $I$
is the region of space occupied by the dielectric material.  This
choice for $I$ is convenient but not unique: the only requirement is
that there is free propagation of light in the external region far
away from the resonator (this allows us to define asymptotic boundary
conditions below). The operators $\mathcal{Q}$ and $\mathcal{P}$
depend on the choice of $I$, but all physical observables turn out to
be independent of this choice.  One easily shows that $\mathcal{P}$
and $\mathcal{Q}$ are projection operators,
\begin{subequations}
\label{eq:prop}
\begin{eqnarray}
\mathcal{P} = \mathcal{P}^\dagger; \hspace*{0.5cm} & 
\mathcal{P}^2 = \mathcal{P}, \\
\mathcal{Q} = \mathcal{Q}^\dagger; \hspace*{0.5cm} & 
\mathcal{Q}^2 = \mathcal{Q} .
\end{eqnarray}
\end{subequations}
Moreover, they are orthogonal, $\mathcal{QP} = \mathcal{PQ} = 0$,
and complete, $\mathcal{Q} + \mathcal{P} = 1$. Therefore, an
arbitrary Hilbert space function ${\bm \phi}$ and the associated
function ${\bf f} = {\bm \phi}/\sqrt{\epsilon}$ may be decomposed
into the projections onto the resonator and channel space
\begin{subequations}
\label{eq:uvdef}
\begin{eqnarray}
{\bm \phi}({\bf r}) & = &  \chi_{-}({\bf r}) {\bm \mu} ({\bf r}) 
+ \chi_{+}({\bf r}) {\bm \nu} ({\bf r}), \\
{\bf f}({\bf r}) & = & \chi_{-}({\bf r}) {\bf u} ({\bf r}) 
+ \chi_{+}({\bf r}) {\bf v} ({\bf r}),
\end{eqnarray}
\end{subequations}
where $\chi_{\mp}$ are the characteristic functions of the resonator
and the channel region, respectively,
\begin{eqnarray}
\chi_{-}({\bf r}) = \int_{{\bf r}^\prime \in I} \! d{\bf r}^\prime 
\delta({\bf r} - {\bf r}^\prime) , \!\quad
\chi_{+}({\bf r}) &=&  1-\chi_{-}({\bf r}).
\end{eqnarray}

Acting on ${\bm \phi}$ with the differential operator $L$ defined in
Eq.~(\ref{eq:hevequniv}), we obtain
\begin{eqnarray}
\label{eq:lqph}
L {\bm \phi} &=& \chi_{-}({\bf r}) L \, {\bm \mu} ({\bf r}) -
\int_{\partial I} \! d^2{\bf r}^\prime K({\bf r}, {\bf r}^\prime)
{\bm \mu} ({\bf r}^\prime) \nonumber \\
&& +\chi_{+}({\bf r}) L \, {\bm \nu}({\bf r}) + \int_{\partial I} \! d^2{\bf
r}^\prime K({\bf r}, {\bf r}^\prime) {\bm \nu} ({\bf r}^\prime) ,
\quad
\end{eqnarray}
where $K({\bf r}, {\bf r}^\prime)$ is a singular differential
operator defined at the boundary
\begin{eqnarray}
\label{eq:defk}
K({\bf r}, {\bf r}^\prime) {\bm \mu}({\bf r}^\prime) & =&   
\left[ \frac{\delta({\bf r}-{\bf r}^\prime) {\bf n}^\prime}
{\sqrt{\epsilon({\bf r})}} \right]  \times  \left[ \nabla^\prime 
\times  \frac{{\bm \mu}({\bf r}^\prime)}{\sqrt{\epsilon
 ({\bf r}^\prime)}} \right] \nonumber \\
& & \!\!\!\!\! + 
\left[ \frac{\nabla \delta({\bf r}-{\bf r}^\prime)}
{\sqrt{\epsilon({\bf r})}} \right] 
\times \left[ {\bf n}^\prime \times  
\frac{{\bm \mu}({\bf r}^\prime)}{\sqrt{\epsilon({\bf
  r}^\prime})}\right]. \quad 
\end{eqnarray}
Here $\nabla^\prime$ denotes a derivative with respect to ${\bf
  r}^\prime$ and ${\bf n}^\prime$ is a unit vector normal to the
boundary. The action of $K({\bf r}, {\bf r}^\prime)$ on ${\bm \nu}
({\bf r}^\prime)$ is defined in a similar fashion, with ${\bm \mu}
({\bf r}^\prime)$ replaced by ${\bm \nu} ({\bf r}^\prime)$.

The first (third) term on the right hand side of Eq.\ 
(\ref{eq:lqph}) contribute only inside (outside) the dielectric. The
second and fourth term are boundary terms. They involve ${\bm \mu}$,
${\bm \nu}$ and their derivatives at the boundary; these functions
must be evaluated in the limit where the boundary is approached from
the cavity and the channel region, respectively. We note that the
boundary terms generally gives rise to singular behavior. As a
result, the action of $L$ usually goes beyond the Hilbert space.
The range of $L$ within the Hilbert space is defined by the
functions for which the singular terms vanish; this happens in
particular for the eigenfunctions of $L$ in Hilbert space.

We now want to decompose the operator $L$ into a resonator, a
channel, and a coupling contribution. However, it is not obvious how
the decomposition can be carried out for the singular boundary
terms. We therefore replace the boundary integrals by integrations
along surfaces arbitrarily close to the boundary but located inside
respectively outside the resonator region. Equation (\ref{eq:lqph})
then becomes
\begin{equation}
\label{eq:Lpro}
L {\bm \phi} = L_{\mathcal{QQ}} {\bm \mu} + L_{\mathcal{QP}}{\bm \nu} + 
L_{\mathcal{PQ}} {\bm \mu} + L_{\mathcal{PP}} {\bm \nu} ,
\end{equation}
where $L_{\mathcal{QQ}}$ and $L_{\mathcal{PP}}$ are the projections of
$L$ onto the resonator and channel space,
\begin{subequations}
\label{eq:lblock1}
\begin{eqnarray}
\label{eq:lsep1}
&&\hspace*{-0.6cm} L_{\mathcal{QQ}} {\bm \mu}=
\chi_{-}({\bf r}) \, L {\bm \mu} ({\bf r}) \nonumber \\
&&
- \int_{\partial I} \! d^2{\bf r}^{\prime}_{-} 
 \bigg[ \frac{\delta({\bf r} - 
{\bf r}^\prime_{-}) {\bf n}^\prime} 
{\sqrt{\epsilon({\bf r})}} \bigg] 
\times \bigg[\nabla^\prime \times \frac{{\bm \mu} ({\bf
r}^\prime_{-})}{\sqrt{\epsilon({\bf r}^\prime_{-})}}\bigg], \quad \quad 
\end{eqnarray}
\begin{eqnarray}
\label{eq:lsep2}
&&\hspace*{-0.6cm} L_{\mathcal{PP}} {\bm \nu}  =
\chi_{+}({\bf r}) \, L {\bm \nu} ({\bf r}) \nonumber \\
&&
+ \int_{\partial I} \! d^2{\bf r}^{\prime}_{+} \bigg[ 
\frac{\nabla \delta({\bf r} - {\bf
r}^\prime_{+})}
{\sqrt{\epsilon({\bf r})}} \bigg]  
\times \bigg[ {\bf n}^\prime  \times 
\frac{{\bm \nu} ({\bf r}^\prime_{+})}
{\sqrt{\epsilon({\bf r}^\prime_{+})} } \bigg], \quad \quad
\end{eqnarray}
\end{subequations}
and $L_{\mathcal{QP}}$ and $L_{\mathcal{PQ}}$ the coupling terms
\begin{subequations}
\label{eq:lblock2}
\begin{eqnarray}
\label{eq:lsep3}
L_{\mathcal{QP}} {\bm \nu}\!\! &=&\!\!\!
+\!\!\int_{\partial I} \! d^2{\bf r}^{\prime}_{-}
\bigg[ \frac{\delta({\bf r} - {\bf
r}^\prime_{-}) {\bf n}^\prime}
{\sqrt{\epsilon({\bf r})}} \bigg]\!\! \times \!\! \bigg[ 
\nabla^\prime \! \times \frac{{\bm \nu} ({\bf
r}^\prime_{+})}{\sqrt{\epsilon({\bf r}^\prime_{+})}} \bigg], 
\quad \quad \, 
\end{eqnarray}
\begin{eqnarray}
\label{eq:lsep4}
L_{\mathcal{PQ}} {\bm \mu} \!\!&=& \!\!- \!\!
\int_{\partial I} \!  d^2{\bf r}^{\prime}_{+}
\bigg[ \frac{\nabla \delta({\bf r} - {\bf
r}^\prime_{+})}{\sqrt{\epsilon({\bf r})}}  \bigg] \! \! 
\times \! \bigg[ {\bf n}^\prime \!\!\times \frac{{\bm \mu} ({\bf
r}^\prime_{-})}{\sqrt{\epsilon({\bf r}^\prime_{-})}}
 \bigg]. \quad \quad \, 
\end{eqnarray} 
\end{subequations}
The shorthands ${\bf r}^\prime_{\mp}$ indicate that the integrals
are to be evaluated in the limit where ${\bf r}^\prime_{-}$ and
${\bf r}^\prime_{+}$ approach the boundary from inside respectively
outside the resonator.  One can easily show that the decomposition
(\ref{eq:Lpro}) preserves the Hermiticity of $L$. Moreover,
$L_{\mathcal{QQ}}$ and $L_{\mathcal{PP}}$ define Hermitean operators
on the Hilbert space functions of the resonator and the channel
region, respectively.

Substitution of Eq.~(\ref{eq:Lpro}) into the eigenmode equation
(\ref{eq:hevequniv}) yields two coupled equations for the
projections of the eigenfunctions onto the resonator and channel
space:
\begin{equation}
\label{eq:heveqsep}
\left( \begin{array}{cc}
       L_{\mathcal{QQ}} &  L_{\mathcal{QP}} \\
       L_{\mathcal{PQ}} &  L_{\mathcal{PP}}
       \end{array} \right)
\left( \begin{array}{c}
       {\bm \mu} (\omega)  \\
       {\bm \nu} (\omega)
       \end{array} \right)
= \frac{\omega^2}{c^2}
\left( \begin{array}{c}
       {\bm \mu} (\omega) \\
       {\bm \nu} (\omega)
       \end{array} \right).
\end{equation}
The condition that the singular terms on the left hand side vanish
yields the two matching conditions
\begin{subequations}
\label{eq:match1}
\begin{eqnarray}
{\bf n} \times \left[ {\bf u}(\omega)-{\bf v}(\omega) \right] & = &
{\bm 0}, \\ {\bf n} \times \left[ \nabla \times {\bf u}(\omega)-
\nabla \times {\bf v}(\omega) \right] & = & {\bm 0},
\end{eqnarray}
\end{subequations}
for all points along the boundary of the resonator region. The gauge
condition $\nabla \cdot [\epsilon {\bf f}(\omega)] =0$ and the
requirement $\nabla \cdot (\nabla \times {\bf f} (\omega))=0$ along
the boundary give the further matching conditions
\begin{subequations}
\label{eq:match3}
\begin{eqnarray}
{\bf n} \cdot \left[ \epsilon {\bf u}(\omega)-\epsilon {\bf
v}(\omega) \right] & = & 0 , \\ {\bf n} \cdot \left[ \nabla \times
{\bf u}(\omega)- \nabla \times {\bf v}(\omega)\right] & = & 0 .
\end{eqnarray}
\end{subequations}
The four matching conditions (\ref{eq:match1}), (\ref{eq:match3})
together with Eq.\ (\ref{eq:ebfield}) realize the well--known
\cite{Jac75} boundary conditions for the electromagnetic field at an
interface in the absence of surface currents or surface charges.

\subsection{Eigenmodes of the resonator and channel region}
\label{sec:eigenmodes}

The operators $L_{\mathcal{QQ}}$ and $L_{\mathcal{PP}}$ are
self--adjoint operators in the Hilbert space of the resonator and
the channel functions, respectively. The eigenfunctions of
$L_{\mathcal{QQ}}$ satisfy the equation
\begin{equation}
\label{eq:eveqqq}
\nabla \times \left[ \nabla \times {\bf u}_{\lambda} ({\bf r}) 
\right] = \frac{\epsilon({\bf r}) \omega^2_{\lambda}}{c^2} 
{\bf u}_{\lambda}({\bf r}).
\end{equation}
From the condition that the singular term in Eq.\ (\ref{eq:lsep1})
vanishes one obtains the boundary condition
\begin{equation}
\label{eq:bound1}
{\bf n} \times \left[ \nabla \times {\bf u}_{\lambda} \right] 
\big|_{\partial I} = {\bm 0} .
\end{equation}
Hence, the tangential component of $\nabla \times {\bf u}_\lambda$
vanishes at the boundary. No boundary condition for the normal
component of $\nabla \times {\bf u}_\lambda$ is required as the
three components of this vector are connected through the gauge
condition. We note that the eigenmodes of the resonator form a
discrete set.

In a similar fashion the eigenmodes of the channel region are found
from Eq.~(\ref{eq:lsep2}). They satisfy the equation
\begin{equation}
\label{eq:eveqpp}
\nabla \times \left[ \nabla
  \times {\bf v}_m(\omega,{\bf r}) \right] = \frac{\epsilon ({\bf
    r}) \omega^2}{c^2} {\bf v}_{m}(\omega,{\bf r}),
\end{equation}
and the condition that the tangential component must vanish at the
boundary,
\begin{equation}
\label{eq:bound2}
{\bf n} \times  {\bf v}_m(\omega) \big|_{\partial I} = {\bm 0} .
\end{equation}
The channel modes form a continuum, labeled by the frequency
$\omega$ and the index $m$ that specifies the asymptotic conditions
at infinity. We note that the resonator modes ${\bf u}_\lambda$ have
support only within the resonator and vanish in the channel region;
vice versa the channel functions ${\bf v}_m (\omega)$ vanish inside
the resonator and take nonzero values only within the channel
region. The resonator and channel modes form complete and
orthonormal basis sets for the resonator and channel region,
respectively. As a result, the projectors $\mathcal{P}$ and
$\mathcal{Q}$ can be represented in terms of these modes,
\begin{subequations}
\label{eq:qpbas}
\begin{eqnarray}
\mathcal{Q} &=& \sum_{\lambda} |{\bm \mu}_{\lambda}\rangle \langle
{\bm \mu}_{\lambda}|, \\ 
\mathcal{P} &=& \sum_m \int \!  d\omega \; |{\bm
\nu}_{m}(\omega)\rangle \langle {\bm \nu}_{m}(\omega)|.
\end{eqnarray}
\end{subequations}
Together with the eigenmode equation (\ref{eq:heveqsep}) this
reduces the eigenmode problem to the well--known problem
\cite{Fan61,Dit00} of a discrete number of states coupled to a
continuum.

We note that the boundary conditions (\ref{eq:bound1}),
(\ref{eq:bound2}) on the resonator and channel modes are a consequence
of our separation of the singular terms in Eq.~(\ref{eq:lqph}). This
separation is by no means unique: Different separations are possible
and generally give rise to different boundary conditions. For example,
the substitution of $\delta( {\bf r}-{\bf r}^\prime)$ by $\delta( {\bf
r}-{\bf r_+}^\prime)$ and of $\nabla \delta( {\bf r}-{\bf r}^\prime)$
by $\nabla \delta( {\bf r}-{\bf r_-}^\prime)$ in Eq.~(\ref{eq:defk}),
leads to a new set of boundary conditions for which the conditions on
the internal and external eigenmodes are just interchanged.  Further
choices are possible subject to the condition that the decomposition
of $L$ is self--adjoint. The freedom in choosing the boundary
conditions is characteristic for the projector technique
\cite{Fes62,Som02}.

It is worth emphasizing that neither the modes ${\bm
\mu}_{\lambda}$ of the closed resonator nor the channel modes ${\bm
\nu}_{m}(\omega)$ represent eigenmodes of the total system. The
latter modes satisfy the matching conditions derived earlier but,
in general, neither of the boundary conditions (\ref{eq:bound1}),
(\ref{eq:bound2}). Still the eigenmodes of the total system may be
expanded in terms of the resonator and channel modes as these modes
form complete basis sets in the respective regions. This fact is a
consequence of the convergence in Hilbert space which {\em does
not} imply pointwise convergence (at the boundary). Consequently,
the matching conditions must not be imposed directly at the
boundary but hold, as usual \cite{Jac75}, in a limiting sense
infinitesimally close to the boundary. A further discussion of the
physical meaning of the resonator and channel modes, as well as
explicit applications of the projector technique to the potential
scattering of matter waves can be found in Ref.\ \cite{Som02}.

The eigenmode equation (\ref{eq:heveqsep}) in full space may now be
solved by standard methods \cite{Fan61,Dit00,Coh92}. For the
projection onto the channel space one has the Lippmann--Schwinger
type solution
\begin{equation}
\label{eq:ppj}
\mathcal{P}|{\bm \phi} (\omega)\rangle = |{\bm \nu}
(\omega)\rangle +
\frac{1}{\frac{\omega^2}{c^2} - L_{\mathcal{PP}}+ i\epsilon} 
L_{\mathcal{PQ}} |{\bm \phi} (\omega)\rangle,
\end{equation}
where the limit $\epsilon \to 0^+$ is implied. Substitution into
the equation for the projection onto the resonator space yields
\begin{equation}
\label{eq:qpj}
\mathcal{Q}|{\bm \phi} (\omega)\rangle = \frac{1}{ 
\frac{\omega^2}{c^2} - L_{\textrm{eff}}(\omega)}
 L_{\mathcal{QP}} |{\bm \nu} (\omega)\rangle,
\end{equation}
where $L_{\textrm{eff}}$ is the non--Hermitean
operator
\begin{equation}
L_{\textrm{eff}} (\omega) \equiv L_{\mathcal{QQ}} + 
L_{\mathcal{QP}} \frac{1}{\frac{\omega^2}{c^2} 
- L_{\mathcal{PP}}+ i\epsilon} L_{\mathcal{PQ}}.
\end{equation}
To simplify notation, we introduce the Green function of the
resonator in the presence of the coupling to the channels
\begin{eqnarray}
G_{\mathcal{QQ}} \left(\frac{\omega^2}{c^2}\right) & = & \frac{1}{
\frac{\omega^2} {c^2} - L_{\textrm{eff}}(\omega)}.
\end{eqnarray}
Combining Eqs.~(\ref{eq:ppj}), (\ref{eq:qpj}) we arrive at an
expression for the eigenstates $|{\bm \phi}(\omega)\rangle$,
\begin{eqnarray}
\label{eq:scs}
&& \hspace*{-0.3cm} |{\bm \phi} (\omega)\rangle = G_{\mathcal{QQ}} 
L_{\mathcal{QP}}  |{\bm \nu} (\omega)\rangle \nonumber \\
&&+ \left[ 1 + \frac{1}{\frac{\omega^2}
{c^2}-L_{\mathcal{PP}}+ i\epsilon} L_{\mathcal{PQ}} 
G_{\mathcal{QQ}} L_{\mathcal{QP}} \right] |{\bm \nu} 
(\omega)\rangle. \quad
\end{eqnarray}
Using the expansion (\ref{eq:qpbas}), this yields an exact
representation of the eigenstates in terms of the resonator and
channel modes
\begin{equation}
\label{eq:univsplit}
|{\bm \phi} (\omega)\rangle = \sum_{\lambda} 
\alpha_{\lambda} (\omega) | {\bm \mu}_{\lambda}\rangle   + 
\int \! d\omega^\prime \; \beta (\omega,\omega^\prime) 
|{\bm \nu} (\omega^\prime)
\rangle  ,
\end{equation}
where the expansion coefficients are given by
\begin{subequations}
\label{eq:exa}
\begin{eqnarray}
\alpha_{\lambda}(\omega) = \langle {\bm \mu}_{\lambda}|
G_{\mathcal{QQ}} L_{\mathcal{QP}} | {\bm \nu} (\omega)
\rangle, \quad \quad
\quad \quad
\end{eqnarray}
\begin{eqnarray} 
&&\beta (\omega,\omega^\prime) = \langle {\bm \nu} (\omega^\prime)|
\bigg[ 1+ \nonumber \\ && \quad \quad \quad \frac{1}{\frac{\omega^2}
{c^2} + i \epsilon -L_{\mathcal{PP}}} L_{\mathcal{PQ}} 
G_{\mathcal{QQ}} L_{\mathcal{QP}} \bigg] |{\bm \nu} 
(\omega)\rangle. \quad \quad
\end{eqnarray}
\end{subequations}
The modes ${\bf f}(\omega)$ of the electromagnetic field
are recovered from ${\bm \phi}(\omega)$ using Eqs.\
(\ref{eq:fguniv}), (\ref{eq:uvdef}):
\begin{equation}
\label{eq:fsep}
{\bf f} (\omega,{\bf r}) = \sum_{\lambda} \alpha_{\lambda} (\omega) 
{\bf u}_{\lambda}({\bf r})  + \int \!  d\omega^\prime \; 
\beta (\omega,\omega^\prime) {\bf v} (\omega^\prime,{\bf r})  .
\end{equation}

\subsection{Field expansions and Hamiltonian}
\label{sec:expans}

The decomposition of the electromagnetic field modes into a
resonator and a channel contribution suggests a quantization scheme
different from the modes--of--the--universe approach discussed in
Sec.\ \ref{sec:normal}. In this section we carry out the field
quantization on the basis of the resonator and channel modes. Our
starting point is the expansion of the vector potential and the
canonical momentum in terms of these modes; combining Eqs.\ 
(\ref{eq:qpexp}), (\ref{eq:fsep}) this expansion takes the form
\begin{subequations}
\label{eq:apsplit}
\begin{eqnarray}
{\bf A}({\bf r},t) &=& c \sum_{\lambda} Q_{\lambda} 
{\bf u}_{\lambda}({\bf r})  + c \int \! d\omega \; Q (\omega) 
{\bf v} (\omega,{\bf r}) , \hspace*{0.7cm} \\ 
{\bm \Pi}({\bf r},t) &=& \frac{1}{c} \sum_{\lambda} 
{\bf u}^{*}_{\lambda}({\bf r}) P_{\lambda} + \frac{1}{c} \int 
\! d\omega \;  {\bf v}^{\dagger}(\omega,{\bf r}) P (\omega) ,
\hspace*{0.7cm}
\end{eqnarray}
\end{subequations}
where we have defined the position operators 
\begin{subequations}
\label{eq:qop}
\begin{eqnarray}
Q_{\lambda} &=& \int \! d\omega \; q(\omega) \alpha_{\lambda} (\omega)
, \\
Q (\omega) &=&  \int \! d\omega^\prime \; q (\omega^\prime) \beta
(\omega,\omega^\prime) ,
\end{eqnarray}
\end{subequations}
and the momentum operators
\begin{subequations}
\label{eq:pop}
\begin{eqnarray}
P_{\lambda} &=& \int \! d\omega \; \alpha^{\dagger}_{\lambda} 
(\omega) p (\omega) , \\
P (\omega) &=& \int \! d\omega^\prime \; \beta^{\dagger}
(\omega,\omega^\prime) p (\omega^\prime) . 
\end{eqnarray}
\end{subequations}
The $Q_\lambda$ and $P_\lambda$ are time--dependent operators that
represent complex amplitudes associated with the resonator field.
Likewise, the operators $Q(\omega)$ and $P(\omega)$ are amplitudes
describing the channel field. The (equal--time) commutation
relations of the various operators are discussed in Appendix
\ref{app:A}. The calculation shows that operators associated with
different subsystems commute.  Moreover, for each subsystem $Q$ and
$P$ behave like the fundamental operators for position and momentum,
respectively.

To discuss the dynamical evolution of the resonator and channel
operators, we must express the field Hamiltonian in terms of these
operators. We use Eqs.\ (\ref{eq:qop}), (\ref{eq:pop}) and the
completeness relation $\mathcal{Q}+\mathcal{P} =1$ to invert the
relation between the operators for the total system and the operator
for the two subsystems
\begin{subequations}
\label{eq:qpsplit}
\begin{eqnarray}
q (\omega) &=& \sum_{\lambda} \alpha^{\dagger}_{\lambda}(\omega)
Q_{\lambda} + \int \! d\omega^\prime \; \beta^{\dagger} 
(\omega,\omega^\prime) Q (\omega^\prime), \quad \\
p (\omega) &=& \sum_{\lambda} P_{\lambda} \alpha_{\lambda}(\omega)  +
\int \! d\omega^\prime \; P (\omega^\prime) \beta 
(\omega^\prime,\omega) . \quad
\end{eqnarray}
\end{subequations}
Substitution into Eq.\ (\ref{eq:hqp}) yields the desired expression
for the field Hamiltonian. Using relations between the expansion
coefficients $\alpha$ and $\beta$ that follow from the completeness
and orthogonality of the modes functions (see Appendix \ref{app:B})
we can write the result in the form
\begin{eqnarray}
\label{eq:aaham}
H &=& \sum_{\lambda} \left[ P^\dagger_{\lambda} P_{\lambda} + 
\omega^2_{\lambda} Q^{\dagger}_{\lambda} Q_{\lambda} \right] \nonumber \\
& & + \sum_m \int \! d\omega \left[ P_m^{\dagger}(\omega) P_m(\omega) 
+ \omega^2 Q_m^{\dagger}(\omega) Q_m(\omega) \right] \quad \nonumber\\
& & + \sum_{\lambda} \sum_m \int \! d\omega \left[ W_{\lambda m}
(\omega) Q^{\dagger}_{\lambda} Q_m(\omega) + \mbox{ h.c.} \right], \quad 
\end{eqnarray}
with $W_{\lambda m}(\omega) = (c^2 /2) \langle {\bm \mu}_{\lambda}|
L |{\bm \nu}_m({\omega})\rangle$. This shows that the operators of
the subsystems do not simply oscillate, as they would if the
subsystems were completely isolated from each other. The origin for
this is the third term on the right hand side of Eq.\ 
(\ref{eq:aaham}) which couples the motion of the resonator and
channel operators. The coupling reflects the fact that the boundary
of the dielectric is not completely reflecting; thus radiation may
leak through the boundary to the external radiation field.

The operators $Q$ and $P$ have a standard representation in terms of
creation and annihilation operators,
\begin{subequations}
\label{eq:qpbres}
\begin{eqnarray}
Q_{\lambda} &=& \left[ \frac{\hbar}{2 \omega_{\lambda}} 
\right]^{\frac{1}{2}} \left[ a_{\lambda} +  \sum_{\lambda^\prime} 
\mathcal{N}_{\lambda \lambda^\prime}^{\dagger} 
a_{\lambda^\prime}^{\dagger}  \right] , \\
P_{\lambda} &=& i \left[ \frac{\omega_{\lambda}}{2\hbar} 
\right]^{\frac{1}{2}} \left[ a_{\lambda}^{\dagger} -  
\sum_{\lambda^\prime} \mathcal{N}_{\lambda \lambda^\prime}
 a_{\lambda^\prime} \right],
\end{eqnarray}
\end{subequations}
where the matrix element $\mathcal{N}_{\lambda \lambda^\prime}$ is
the overlap integral
\begin{eqnarray}
\label{eq:udis}
\mathcal{N}_{\lambda \lambda^\prime} = \int \! d{\bf r} \;
\epsilon({\bf r}){\bf u}_{\lambda}
({\bf r}) \cdot {\bf u}_{\lambda^\prime}({\bf r}) .
\end{eqnarray}
The operators $a_\lambda$ and $a_{\lambda^\dagger}$ obey the
canonical commutation relations
\begin{eqnarray}
[ a_{\lambda}, a_{\lambda^\prime}^\dagger ]  =  \delta_{\lambda 
\lambda^\prime}, \qquad
{[ a_\lambda , a_{\lambda^\prime} ]}  =  0.
\end{eqnarray}
In a similar fashion one derives the representation of the channel
operators $Q_m(\omega)$, $P_m(\omega)$ in terms of a continuous set
of creation and annihilation operators $b_m^\dagger (\omega)$, $b_m
(\omega)$. Substituting these representations into the Hamiltonian
(\ref{eq:aaham}) and using the symmetry and unitarity of the overlap
matrices, one arrives at the Hamiltonian
\begin{eqnarray}
\label{eq:hlast}
H &=& \sum_{\lambda} \hbar \omega_{\lambda} \, a^{\dagger}_{\lambda}
a_{\lambda} + \sum_m \int \! d\omega \; \hbar\omega \, b_m^{\dagger}
(\omega) b_m(\omega) \nonumber \\
& & + \hbar \sum_{\lambda} \sum_m \int \! d\omega \bigg[ 
\mathcal{W}_{\lambda m}(\omega) \, a^{\dagger}_{\lambda} 
b_{m}(\omega) \nonumber \\
& & \hspace*{1.8cm}  +\mathcal{V}_{\lambda
    m}(\omega) \, a_{\lambda} b_{m}(\omega) + \mbox{h.c.} \bigg], 
\end{eqnarray}
where we have omitted an irrelevant zero point contribution. The
coupling matrix elements are given by
\begin{subequations}
\begin{eqnarray}
\mathcal{W}_{\lambda m}(\omega) &=& \frac{c^2}{2 \hbar
  \sqrt{\omega_\lambda \omega}}  \langle {\bm \mu}_{\lambda} | 
L_{\mathcal{QP}} |{\bm \nu}_m(\omega)
\rangle , \\
\label{eq:coupv}
\mathcal{V}_{\lambda m}(\omega) &=& \frac{c^2}{2 \hbar 
\sqrt{\omega_\lambda \omega}} \langle {\bm \mu}_{\lambda}^{*} 
| L_{\mathcal{QP}} |{\bm \nu}_m(\omega)\rangle .
\end{eqnarray}
\end{subequations}
The notation $\langle {\bm \mu}_{\lambda}^{*} |$ means $\langle {\bm
  \mu}_{\lambda}^{*} |{\bf r} \rangle \equiv {\bm \mu} ({\bf r})$.
Finally, substituting the representation (\ref{eq:qpbres}) into
Eq.~(\ref{eq:apsplit}) we find the expansion of the intracavity
field
\begin{subequations}
\label{eq:afin}
\begin{eqnarray}
{\bf A}({\bf r},t) & = & c \sum_\lambda \left[\frac{\hbar}{2
\omega_\lambda}  \right]^{1/2} [ a_\lambda {\bf u}_\lambda 
({\bf r})   + a_\lambda^\dagger 
 {\bf u}^*_\lambda ({\bf r})   ] , \quad \quad \\
{\bm \Pi} ({\bf r},t) & = & \! \! -\frac{i}{c} 
\sum_\lambda \left[\frac{\hbar 
\omega_\lambda}{2}  \right]^{1/2} \! \! [ a_\lambda  
{\bf u}_\lambda ({\bf r})-a_\lambda^\dagger  
{\bf u}^*_\lambda ({\bf r}) ]. \quad \quad
\end{eqnarray}
\end{subequations}
The Hamiltonian (\ref{eq:hlast}) and the field expansions
(\ref{eq:afin}) are the key results of the quantization procedure.
The field expansions of the open resonator reduce precisely to the
standard expressions known from closed resonators. However, the
field {\em dynamics} is fundamentally different as shown below. We
note that the resonator modes are coupled to the exter\-nal
radiation field via both resonant ($a^\dagger b$, $b^\dagger a$) and
non--resonant ($ab$, $a^\dagger b^\dagger$) terms. The non-resonant
terms become important in the case of overdamping (when the mode
widths are comparable to the optical frequencies) \cite{Haa85}.  In
most cases of interest, the widths are much smaller than the
relevant frequencies; then the rotating--wave approximation can be
made, which amounts to keeping only the resonant terms in the
Hamiltonian.  In this approximation, the Hamiltonian
(\ref{eq:hlast}) reduces to the well--known system--and--bath
Hamiltonian \cite{Sen59,Gar00} of quantum optics.  It has been
argued, that this Hamiltonian is valid only for good cavities with
spectrally well--separated modes. Our derivation shows that such
pessimism is inappropriate: the system--and--bath Hamiltonian does
describe the dynamics of overlapping modes, provided the broadening
of these modes is much smaller than their frequency (so that
non--resonant terms can be neglected).

\section{Field dynamics}
\label{sec:linear}

Measurements on optical cavities are typically done with detectors
located in the external region outside the cavity. The detectors
therefore measure the external field, and an input--output theory is
required to relate the evolution of the external field to the
dynamics of the system of interest. In this section we apply the
input--output formalism of Gardiner and Collet \cite{Gar00,Wal94} to
the dynamics generated by the system--and--bath Hamiltonian. We show
that there is a linear relation $b^{\textrm{out}}(\omega)=S(\omega)
b^{\textrm{in}}(\omega)$ between the cavity input and output field,
involving the scattering matrix $S(\omega)$ of the cavity. Our
derivation of $S(\omega)$ differs from most applications of the
input--output formalism in two respects. First, we do not impose the
Markov approximation and do not require that the coupling amplitudes
between cavity and channel modes are frequency independent. Second,
we do not restrict ourselves to essentially one--dimensional
scattering, but consider the more general situation with multiple
input and output channels and a non--trivial cavity dynamics. In
particular, we will address the case of chaotic scattering, i.e.\ 
the case when the propagation of light in the cavity becomes chaotic
due to random fluctuations of the refractive index or due to
scattering at irregularly shaped mirrors.

\subsection{Input--output relation}
\label{sec:inout}
The starting point of the input--output theory \cite{Gar00,Wal94}
are the equations of motion for the annihilation operators of the
intracavity and channel field modes. These equations take a
particular simple form when the damping rates are much smaller
than the frequencies of interest. This permits us to use the
rotating--wave approximation, in which the Hamiltonian
(\ref{eq:hlast}) reduces to the system--and--bath Hamiltonian
\begin{eqnarray}
\label{eq:hsb}
H_{\textrm{SB}} &=& \sum_{\lambda} \hbar \omega_{\lambda} \, 
a^{\dagger}_{\lambda}
a_{\lambda} + \sum_m \int \! d\omega \; \hbar\omega \, b_m^{\dagger}
(\omega) b_m(\omega) \nonumber \\
&& \hspace*{-0.6cm} + \hbar \sum_{\lambda} \sum_m \int 
\! d\omega \left[ \mathcal{W}_{\lambda m}(\omega) \,
  a^{\dagger}_{\lambda} b_{m}(\omega) + \mbox{ h.c. }\right] . 
\quad \quad
\end{eqnarray}
Here we have extended the range of the frequency integrals from
$-\infty$ to $\infty$ consistent with the rotating--wave
approximation \cite{Gar00}. The Heisenberg equations of motion for
the internal operators $a_\lambda$ and the channel operators
$b_m(\omega)$ are given by
\begin{eqnarray}
\label{eq:aieqm}
\dot{a}_{\lambda} &=& -i \omega_\lambda a_{\lambda} - i \sum_m 
\int \! d\omega  \; \mathcal{W}_{\lambda
  m}(\omega) b_m(\omega), \quad \\
\label{eq:aoeqm}
\dot{b}_m(\omega) &=& -i\omega b_m(\omega) - i \sum_{\lambda}
\mathcal{W}^{*}_{\lambda m}(\omega) a_{\lambda}. \quad
\end{eqnarray}
Integration of Eq.\ (\ref{eq:aoeqm}) starting from some initial
time $t_0<t$ yields
\begin{eqnarray}
\label{eq:aineqm}
{b}_m(\omega,t) &=& e^{-i\omega (t-t_0)} b_m(\omega,t_0) \nonumber \\
& & \hspace*{-0.3cm} - i \sum_{\lambda} \mathcal{W}^{*}_{\lambda
m}(\omega) \int\limits^{t}_{t_0} \!  dt^{\prime} \;
e^{-i\omega(t-t^{\prime})} a_{\lambda}(t^{\prime}), \ \ \
\end{eqnarray}
where $b_m(\omega,t_0)$ denotes the channel operator $b_m(\omega)$
at time $t_0$. In an analogous fashion, one can express $b_m
(\omega, t)$ in terms of the channel operators at the final time
$t_1>t$,
\begin{eqnarray}
\label{eq:aouteqm}
{b}_m(\omega,t) &=& e^{-i\omega (t-t_1)} b_m(\omega,t_1) \nonumber
\\ && \hspace*{-0.3cm} + i \sum_{\lambda} \mathcal{W}^{*}_{\lambda
m}(\omega) \int\limits^{t_1}_{t} \!  dt^{\prime} \;
e^{-i\omega(t-t^{\prime})} a_{\lambda}(t^{\prime}). \ \ \
\end{eqnarray}
Eventually, we are interested in the limits $t_0\longrightarrow
-\infty$ and $t_1\longrightarrow \infty$. Substracting
Eq.~(\ref{eq:aineqm}) from Eq.~(\ref{eq:aouteqm}) and integrating
the result over frequency, we obtain the input--output relation in
the time domain,
\begin{eqnarray}
\label{eq:in-outt}
& & b^{\textrm{out}}_m(t) - b^{\textrm{in}}_m(t) 
= \nonumber \\
& & \ \ -
\frac{i}{2\pi} \sum_{\lambda} \int \! d\omega \; 
\mathcal{W}^{*}_{\lambda m}(\omega) \int\limits^{t_1}_{t_0} 
\! dt^{\prime} \;
e^{-i\omega(t-t^{\prime})} a_{\lambda}(t^{\prime}). \ \ \
\end{eqnarray}
Equation (\ref{eq:in-outt}) relates the cavity operators in the time
interval $t_0 < t < t_1$ to the input and output field operators
\begin{subequations}
\label{eq:aout}
\begin{eqnarray}
b^{\textrm{out}}_m(t) & \equiv & \frac{1}{2\pi}
\int  \! d\omega \;e^{-i\omega (t-t_1)} b_m(\omega,t_1), \\
b^{\textrm{in}}_m(t) & \equiv & \frac{1}{2\pi} 
\int \! d\omega \;e^{-i\omega (t-t_0)} b_m(\omega,t_0).
\end{eqnarray}
\end{subequations}
Fourier transformation of Eq.~(\ref{eq:in-outt}) yields the
input--output relation in the frequency domain. In the asymptotic
limit $t_0 \to -\infty$, $t_1 \to \infty$ the result takes the
form
\begin{eqnarray}
\label{eq:in-outw}
b^{\textrm{out}} (\omega) - b^{\textrm{in}} (\omega) = - i 
\mathcal{W}^\dagger (\omega) a(\omega),
\end{eqnarray}
where we combined the input and output operators at frequency
$\omega$ to $M$--component vectors.  The coupling amplitudes
$\mathcal{W}_{\lambda m}$ form an $N \times M$ coupling matrix
$\mathcal{W}$, and the cavity mode annihilation operators an
$N$--component vector. The finite number $N$ of cavity modes is
artificial, and will eventually be taken to infinity. We note that
the input and output operators are simply related to the channel
mode operators at time $t_0$ and $t_1$, respectively,
$b^{\textrm{in}} (\omega) = e^{i\omega t_0} b (\omega,t_0)$ and
$b^{\textrm{out}} (\omega) = e^{i\omega t_1} b (\omega,t_1)$.

\subsection{S--matrix}
\label{sec:smatrix}

For linear systems one can eliminate the cavity modes from the
equations of motion to derive a linear relation between the input
and output field. Substitution of Eq.\ (\ref{eq:aineqm}) into the
equations of motion (\ref{eq:aieqm}) for the cavity modes yields a
set of linear differential equations which is readily solved by
Fourier transformation. The result is
\begin{equation}
\label{eq:alin}
a(\omega) = 2\pi D^{-1}(\omega) \mathcal{W}(\omega)
b^{\textrm{in}}(\omega),
\end{equation}
where $D$ is the $N \times N$ matrix with the matrix elements
\begin{eqnarray}
&&D_{\lambda \lambda^{\prime}}(\omega) = (\omega - \omega_{\lambda}) 
\delta_{\lambda \lambda^{\prime}} \nonumber \\
&&+ \int\limits^{t}_{-\infty}
dt^{\prime} \int \! d\omega^{\prime} e^{-i(t-t^{\prime}) (\omega^{\prime}
 - \omega -i\epsilon)} [\mathcal{W}(\omega) \mathcal{W}^\dagger 
(\omega)]_{\lambda \lambda^\prime}. \quad \quad
\end{eqnarray}
We have taken the limit $t_0 \longrightarrow -\infty$ and introduced
the positive infinitesimal $\epsilon$ to regularize the integral.
Note that the positive sign of $\epsilon$ is a consequence of the
fact that $a(\omega)$ is expressed in terms of the channel operators
in the remote past $t_0 \to -\infty$. The integral can be evaluated
using the identity
\begin{equation}
\lim_{\epsilon \rightarrow 0^+} \frac{1}{(\omega^\prime -
  \omega) - i\epsilon} = \mathcal{P} \left( \frac{1}{\omega^\prime -
    \omega} \right) + i\pi \delta(\omega^\prime -\omega).
\end{equation}
The result is
\begin{eqnarray}
\label{eq:d}
D_{\lambda \lambda^{\prime}}(\omega) & = & (\omega - \omega_\lambda) 
\delta_{\lambda \lambda^{\prime}} \nonumber \\
&& + \Delta_{\lambda \lambda^{\prime}} (\omega) + i\pi 
\left[ \mathcal{W}(\omega) \mathcal{W}^{\dagger}(\omega) 
\right]_{\lambda \lambda^{\prime}}, \quad
\end{eqnarray}
where $\Delta_{\lambda \lambda^{\prime}}$ is the principal value
integral
\begin{equation}
\Delta_{\lambda \lambda^{\prime}}(\omega) = \sum_m \mathcal{P} \int \!
d\omega^{\prime} \frac{\mathcal{W}_{\lambda m}(\omega^{\prime}) 
\mathcal{W}^{*}_{m \lambda}(\omega^{\prime})}{\omega^{\prime} - \omega}.
\end{equation}
Combining Eqs.\ (\ref{eq:in-outw}), (\ref{eq:alin}) we can eliminate
the internal operators from the input--output relation. This yields
a linear relation between the incoming and the outgoing field,
\begin{equation}
\label{eq:slinear}
b^{\textrm{out}}(\omega) = S(\omega) b^{\textrm{in}}(\omega), 
\end{equation}
where $S$ is the $M \times M$ scattering matrix,
\begin{equation}
\label{eq:weids}
S(\omega) =  \openone - 2\pi i \mathcal{W}^{\dagger}(\omega) 
D^{-1}(\omega) \mathcal{W}(\omega).
\end{equation}
This representation of the scattering matrix is well--known in nuclear
and condensed matter physics \cite{Mah69,Bee97,Dit00}. It is a
generalization of the usual Breit--Wigner result for a single
resonance to the scattering in the presence of $N$ resonances. Using
the commutation relations $[b_n(\omega),b_m^\dagger (\omega^\prime)] =
\delta_{nm} \, \delta (\omega - \omega^\prime)$ for both $b =
b^{\textrm{in}}$ and $b = b^{\textrm{out}}$, one can easily show that
$S$ is unitary, $S S^\dagger = S^\dagger S =\openone$.  The S--matrix
describes scattering both in the limit of isolated resonances and in
the regime of overlapping resonances. In the context of quantum
optics, the regime of isolated resonances corresponds to the
weak-damping regime, where all matrix elements of $\mathcal{W}
\mathcal{W}^\dagger$ are much smaller than the mean frequency spacing
of the resonator modes. The opposite regime of overlapping resonances
is realized when the damping rates exceed the mean frequency spacing.

According to Eqs.\ (\ref{eq:slinear}), (\ref{eq:weids}) the field
dynamics is governed by the {\em resonances} of the open cavity. The
resonances are the complex poles of the S--matrix. They are the
solutions of the equation $\textrm{det} D(\omega) =0$; from Eq.\
(\ref{eq:d}) they represent the complex eigenvalues of the internal
field dynamics in the presence of damping inflicted by the coupling to
the external radiation field. We note that the resonances determine
the field dynamics even though the underlying field quantization is
formulated in terms of closed--cavity eigenmodes (with boundary
conditions as discussed in Sec.\ \ref{sec:eigenmodes}).

Equation (\ref{eq:weids}) has found widespread application in the
random matrix theory of scattering \cite{Bee97,Dit00}. It is the
starting point for the so-called Hamiltonian approach to chaotic
scattering. This approach assumes that the {\em Hamiltonian} of a
closed chaotic resonator can be represented by a random matrix drawn
from a Gaussian ensemble of random matrix theory. The eigenvalues of
the internal Hamiltonian show level repulsion and universal
statistical properties. The statistics of the scattering matrix is
derived from the distribution of the internal Hamiltonian using
Eq.~(\ref{eq:weids}).  An alternative approach to chaotic scattering
is the random S--matrix approach in which one directly models the
statistical properties of $S$ without introduction of a
Hamiltonian. Based on the latter approach, Beenakker and coworkers
\cite{Bee98,Mis00} recently computed the noise properties of
disordered and chaotic optical resonators. The Hamiltonian and the
S--matrix approach to chaotic scattering are known to be equivalent.
However, the Hamiltonian approach has the advantage that one can
include the interaction with an atomic medium on a microscopic
level. An example is given in Sec.~\ref{sec:spont}.

\subsection{Linear absorbing or amplifying medium}

The presence of an absorbing or amplifying medium within the cavity
leads to additional noise and modifies the input--output relation.
Phenomenologically, the interaction with linear media can be modeled
\cite{Gar00,Bee98} by coupling the cavity modes to additional baths.
An absorbing medium is described by a thermal bath of harmonic
oscillators while an amplifying medium may be represented by a bath
of inverted harmonic oscillators at a negative temperature $-T$. The
total Hamiltonian is then given by
\begin{eqnarray}
\label{eq:hmb}
H &=& H_{\textrm{SB}} + H_{\textrm{abs}}+ H_{\textrm{amp}},
\end{eqnarray}
where $H_{\textrm{SB}}$ is the system--and--bath Hamiltonian
(\ref{eq:hsb}) while $H_{\textrm{abs}}$ and $H_{\textrm{amp}}$
represent the absorbing and amplifying bath,
\begin{subequations}
\label{eq:hamblock}
\begin{eqnarray}
& &H_{\textrm{abs}} =  \sum_l \int \! d\omega \;
\hbar\omega \, c^{\dagger}_{l}(\omega) c_{l}(\omega) \nonumber \\
&&\quad  + \hbar \sum_{\lambda} \sum_l \int \! d\omega \left[
    \kappa_{\lambda l}(\omega) \, a^{\dagger}_{\lambda}
    c_{l}(\omega) + \mbox{h.c.}\right], \quad \quad 
\end{eqnarray}
\begin{eqnarray}
& &H_{\textrm{amp}} = - \sum_k \int 
\! d\omega \; \hbar\omega \, d^{\dagger}_{k}(\omega) 
d_{k}(\omega) \nonumber \\
&&\quad + \hbar \sum_\lambda \sum_k \int \! d\omega 
\left[ \gamma_{\lambda k}(\omega) \, a^{\dagger}_{\lambda} 
d_{k}(\omega) + \mbox{ h.c. }\right]. \quad \quad
\end{eqnarray}
\end{subequations}
The operators $c_l$, $c_l^\dagger$ obey the canonical commutation
relations
\begin{equation}
\label{eq:cab}
[c_l(\omega),c_{l^{\prime}}^{\dagger}(\omega^\prime)] = \delta_{l
  l^\prime}\,\delta(\omega - \omega^\prime),
\end{equation}
and account for thermal emission within the absorbing medium. The
operators $d_k$ and $d_{k^\prime}^\dagger$ represent the amplifying
medium and have the commutation relations \cite{Gar00}
\begin{equation}
\label{eq:dam}
[d_k(\omega),d_{k^{\prime}}^{\dagger}(\omega^\prime)] = -\delta_{k
  k^\prime}\,\delta(\omega - \omega^\prime).
\end{equation}
As the Hamiltonian (\ref{eq:hmb}) gives rise to linear equations of
motion, we can compute the cavity output field using Fourier
transformation. The calculation proceeds along the lines of the
calculation presented in Sec.~\ref{sec:smatrix}. The result is
\begin{equation}
\label{eq:slinm}
b^{\textrm{out}}(\omega) = S(\omega) b^{\textrm{in}}(\omega) + 
U(\omega) c^{\textrm{in}} (\omega) + 
  V(\omega) d^{\textrm{in}}(\omega),
\end{equation}
where $c^{\textrm{in}}$ and $d^{\textrm{in}}$ represent the input
noise of the absorbing and amplifying bath. Both are integrals over
bath operators at the initial time $t_0$,
\begin{subequations}
\label{eq:cin}
\begin{eqnarray}
c^{\textrm{in}}_l(t) & \equiv & \frac{1}{2\pi}
\int \! d\omega \;e^{-i\omega (t-t_0)} c_l(\omega,t_0), \\
d^{\textrm{in}}_k(t) & \equiv & \frac{1}{2\pi} 
\int \! d\omega \;e^{-i\omega (t-t_0)} d_k(\omega,t_0).
\end{eqnarray}
\end{subequations}
The matrices $U$ and $V$ are given by
\begin{subequations}
\begin{eqnarray}
U(\omega) &=&  - 2\pi i \mathcal{W}^{\dagger}(\omega)
    D^{-1}(\omega) \mathcal{K}(\omega),  \\
V(\omega) &=&  - 2\pi i \mathcal{W}^{\dagger}(\omega)
    D^{-1}(\omega) \Gamma(\omega) ,
\end{eqnarray}
\end{subequations}
where the $N\times L$ matrix $\mathcal{K}$ and the $N\times K$ matrix
$\Gamma$ comprise the coupling amplitudes $\kappa_{\lambda l}$ and
$\gamma_{\lambda k}$, respectively. In the presence of the absorbing
and amplifying baths, the elements of the $N \times N$ matrix
$D(\omega)$ have the form
\begin{subequations}
\label{eq:dl}
\begin{eqnarray}
D_{\lambda \lambda^{\prime}}(\omega) & = & (\omega - \omega_\lambda) 
\delta_{\lambda \lambda^{\prime}} + \Delta_{\lambda
  \lambda^{\prime}} (\omega) + i\pi \Sigma_{\lambda
  \lambda^{\prime}} (\omega), \quad \quad \\
\Delta_{\lambda \lambda^{\prime}}(\omega) & = & \mathcal{P} \int \!
d\omega^{\prime} \frac{ \Sigma_{\lambda \lambda^\prime} (\omega^\prime)} 
{\omega^{\prime} - \omega},
\end{eqnarray}
\end{subequations}
where $\Sigma$ is the matrix
\begin{eqnarray}
\Sigma(\omega)  = \mathcal{W}(\omega)\mathcal{W}^{\dagger}
(\omega) + \mathcal{K}(\omega)\mathcal{K}^{\dagger}(\omega) - \Gamma
(\omega) \Gamma^{\dagger}(\omega) .
\end{eqnarray}
Using Eq.\ (\ref{eq:slinm}) and the commutation relations for the
output and input noise operators, one obtains the relation 
\begin{equation}
\label{eq:nonuni}
U U^{\dagger} - V V^{\dagger} = \openone - S S^{\dagger},  
\end{equation}
that was first derived by Beenakker \cite{Bee98} using a scattering
approach to field quantization. We note that the matrix $\openone -S
S^\dagger$ is positive definite in an absorbing medium ($V =0$) and
negative definite in an amplifying medium ($U=0$). The relations
(\ref{eq:slinm}) and (\ref{eq:nonuni}) are important as they relate
the intensity of the output field to the amplitudes of the input
field and the scattering matrix of the cavity. The statistical
properties of the scattering matrix are known from random matrix
theory. This allows \cite{Bee98} to compute moments or even the full
distribution of the output field intensity from linear random media.

\subsection{Langevin equations for the internal modes}

It is frequently impractical or impossible to eliminate the cavity
modes from the equations of motion. This happens, for example, when
the cavity field is coupled to strongly pumped atoms. Then the
dynamics of the total system comprising the field and the atoms
becomes nonlinear, and the S--matrix approach of the previous
section cannot be applied. A standard method for tackling
interaction problems is to solve the equations of motion for the
internal modes. These equations are quantum Langevin equations in
which the coupling to the external radiation field gives rise to
damping and noise.

We derive the Langevin equations in the Markov approximation. In
particular, we will assume that the coupling amplitudes
$\mathcal{W}_{\lambda m}(\omega)$ are independent of frequency
\cite{Fyo97} over a sufficiently large frequency band centered
around the frequency $\omega_0$ of interest (in a laser $\omega_0$
is the atomic transition frequency).  Substituting Eq.\ 
(\ref{eq:aineqm}) into the equation of motion (\ref{eq:aieqm}), and
performing the frequency integral, we obtain the Langevin equations
\begin{eqnarray}
\label{eq:mot}
\dot{a}_\lambda (t) =   -i \omega_\lambda a_\lambda (t)- \! \pi \!
\sum_{\lambda^\prime} [\mathcal{W}\mathcal{W}^\dagger]_{\lambda 
\lambda^\prime}
a_{\lambda^\prime} (t) + F_\lambda (t) , \quad
\end{eqnarray}
where $F_\lambda (t)$ is the noise operator
\begin{eqnarray}
\label{eq:in} 
F_\lambda (t) = -i \int d \omega 
e^{-i \omega (t -t_0)} \sum_m \mathcal{W}_{\lambda m} b_m(\omega,t_0) .
\end{eqnarray}
The Eqs.~(\ref{eq:mot}) generalize the Langevin equation for a single
cavity mode \cite{Wal94} to the case of many modes\cite{Bar99}. We
note that the resulting equations differ from the
independent--oscillator equations of standard laser theory
\cite{Hak84} in two respects: First, the mode operators $a_\lambda$
are coupled by the damping matrix $\mathcal{W}\mathcal{W}^\dagger$;
second, the noise operators $F_\lambda$ are correlated, $\langle
F^\dagger_\lambda F_{\lambda^\prime} \rangle \neq \delta_{\lambda
\lambda^\prime}$, as different modes couple to the same external
channels (the expectation value is defined with respect to the channel
oscillators at time $t_0$). The mode coupling by both damping and
noise can be understood as a consequence of the
fluctuation-dissipation theorem.

The origin of the deviations from the independent oscillator
dynamics may be understood in the limiting case of weak damping.
This is the regime where all matrix elements of $\mathcal{W}
\mathcal{W}^\dagger$ are much smaller than the resonator mode
spacing $\Delta \omega$.  This regime is realized in dielectrics
that strongly confine light due to a large mismatch in the
refractive index.  To leading order in $\mathcal{W}
\mathcal{W}^\dagger / \Delta \omega$ only diagonal elements
contribute to the damping matrix, and Eq.~(\ref{eq:mot}) reduces to
the standard equation of motion for independent oscillators. This
shows that the independent--oscillator dynamics is a limiting case
of the true mode dynamics in the regime of weak damping. Coupled
equations of motion are found when the damping rate becomes of the
order or larger than the mean frequency spacing of the internal
modes.

According to the universality hypothesis of chaotic scattering, the
internal Hamiltonian of chaotic resonators can be represented by a
random matrix from the Gaussian orthogonal ensemble of random-matrix
theory \cite{Bibel}. The eigenvalues $\omega_\lambda$ display level
repulsion and universal statistical properties. From Eq.\
(\ref{eq:mot}), the mode dynamics of open chaotic resonators is not
only determined by the eigenvalues of the internal Hamiltonian but
also by the coupling strength to the external radiation field.
Therefore, the spectrum of such resonators is governed by a {\em
non--Hermitean} random matrix.  We thus encounter an interesting
connection between the spectral properties of chaotic optical
resonators and non--Hermitean random matrices
\cite{Sok89,Fyo97,Cha97}.

\section{Spontaneous emission}
\label{sec:spont}
In the preceding sections have been concerned with the quantum
properties of the electromagnetic field in open resonators. We now
address the interaction of the radiation field with atoms. As a
simple but nontrivial problem we consider the spontaneous emission
of a two--level atom inside a cavity. This problem has attracted
considerable interest \cite{Coh92}; and it was found that the cavity
may drastically modify the rate of spontaneous emission from its
value in free space. The reason for the effect is the modification
due to the cavity of the local density of modes at the position of
the atom.  Most investigations of the spontaneous emission rate
assumed cavities of regular shape, but recently \cite{Mis97,Lam99}
also unstable and chaotic cavities were addressed.  We show below
that our system--and--bath Hamiltonian reproduces the standard
result for the atomic decay rate within the Wigner--Weisskopf
approximation. We express the result in terms of left and right
eigenmodes of a non--Hermitean matrix and demonstrate that for
chaotic resonators a statistical analysis of the decay rate is
possible using random matrix theory.

We consider a single two-level atom with transition frequency
$\omega_0$ located at the position ${\bf r}_0$ inside an open
cavity. The cavity is empty and defined by external mirrors of
arbitrary shape. The coupling between the atom and the field is
described in the dipole approximation; the dipole strength of the
atomic transition is given by ${\bf d} = \langle 0|e {\bf r}|1
\rangle$, where $e$ is the elementary charge and $|1\rangle$,
$|0\rangle$ the excited state and the ground state of the atom. It
is convenient to introduce the lowering and raising operators
$\sigma = |0\rangle\langle 1|$ and $\sigma^{\dagger} =
|1\rangle\langle 0|$, respectively.

In the rotating wave approximation the total Hamiltonian for the
field and the atom has the form
\begin{equation}
\label{eq:at-fham}
H = H_{\textrm{SB}} + \hbar\omega_0 \sigma^{\dagger}\sigma +
\sum_{\lambda} \left[ g_{\lambda}\, a_\lambda \sigma^{\dagger} + 
\mbox{h.c.} \right].
\end{equation} 
The first term on the right hand side is the system--and--bath
Hamiltonian (\ref{eq:hsb}), the second term represents the free
Hamiltonian of the atom and the last terms account for the coupling
between the atom and the cavity modes with the coupling amplitudes
\begin{equation}
g_{\lambda} = -i
\left(\frac{\hbar\omega_{\lambda}}{2}\right)^{\frac{1}{2}}
 {\bf d}\cdot{\bf  u}_{\lambda}({\bf r}_0).
\end{equation}
We note that the rotating wave approximation enters the Hamiltonian
(\ref{eq:at-fham}) in a twofold way: First we neglected rapidly
oscillating non--resonant terms in the atom--field interaction.
Second, we omitted the non--resonant terms in the field Hamiltonian,
assuming that the decay rate of the cavity modes is much smaller
than the frequencies of interest. This second approximation is
convenient but not essential for the following calculation.  The
modifications that arise when the second approximation is dropped are
summarized in Appendix \ref{app:C}.

To compute the spontaneous emission rate we follow the standard
Wigner--Weisskopf procedure. We assume that initially the atom is in
the excited state while there is no photon in the radiation field.
Hence, the state of the total system at time $t=0$ is given by
$|1,{\textrm{vac}}\rangle$, where vac represents the vacuum state of
the electromagnetic field. Since the Hamiltonian (\ref{eq:at-fham})
conserves the total number of atom and field excitations, the
time-dependent solution of the Schr\"odinger equation can be written
in the form
\begin{eqnarray}
|\Phi (t)\rangle &=& c(t) |1,{\textrm{vac}}\rangle + \sum_{\lambda} 
c_{\lambda}(t) |0,1_{\lambda}\rangle \nonumber \\
&&+ \sum_m \int \! 
d\omega \, c_m(\omega,t) |0,1_{m}(\omega)\rangle ,
\end{eqnarray}
where $c_{\lambda}(t)$ and $c_m(\omega,t)$ are, respectively, the
probability amplitude to find a single photon in the cavity mode
$\lambda$ and in channel $m$ with frequency $\omega$. The time
evolution of the amplitudes $c(t)$, $c_{\lambda}(t)$ and
$c_m(\omega,t)$ follows from the Schr\"odinger equation; an exact
solution can be obtained using Laplace transformation. However,
within the framework of the Wigner--Weisskopf approximation $c(t)$
decays exponentially
\begin{equation}
c(t)=\exp \left[- i(\omega_0 +\delta \omega_0)t -\frac{\gamma}{2}t
\right]
c(0),
\end{equation}
where $\delta \omega_0$ is a frequency shift and $\gamma$ the decay
rate of the intensity $|c(t)|^2$. We note that an exponential decay
is only found if the local density of modes is smooth on the scale
of the atomic decay rate. The decay rate is given by
\begin{equation}
\label{eq:gb}
\gamma = \lim_{\epsilon\to 0} {\textrm{Re}} \left[\frac{2}{\hbar^2}
 \sum_{ij} d_i d^{*}_j C_{ij}(\omega_0+i \epsilon)\right],
\end{equation}
where $i$, $j$ label the components of the dipole matrix element.
Here $C_{ij}(\omega_0+i \epsilon)$ is the Fourier transform of the
two time correlation function of the electric field
\begin{equation}
\label{eq:clap}
C_{ij}(t-t^{\prime}) \equiv \Theta (t-t^\prime) 
\left\langle E_i^+({\bf r}_0,t) E_j^-({\bf r}_0,t^{\prime}) 
\right\rangle_{\mbox{\scriptsize vac}},
\end{equation}
$E^{\pm}$ denote the positive and negative frequency part of the
electric field, $\Theta(t-t^\prime)$ is the step function, and the
average $\langle \cdots \rangle_{\textrm{vac}}$ is the
quantum average over the initial state of the field.

The electric field is connected with the canonical momentum field
through the relation ${\bf E}({\bf r},t) = -c {\bm \Pi} ({\bf
  r},t)$. Inside the cavity both can be expanded in terms of the
cavity modes using Eq.\ (\ref{eq:afin}).  Substitution into
Eq.~(\ref{eq:clap}) reduces the field correlation function to a sum
over the Green functions of the cavity modes,
\begin{subequations}
\begin{eqnarray}
\label{eq:2tcorr}
C_{ij}(\tau) & = & i\sum_{\lambda
  \lambda^{\prime}} \frac{\hbar \sqrt{\omega_{\lambda}
  \omega_{\lambda^{\prime}}}}{2} u_{\lambda i}({\bf r}_0) 
  u^{*}_{\lambda^\prime j}({\bf r}_0) G_{\lambda \lambda^{\prime}}
 (\tau), \quad \quad \\
\label{eq:green1}
G_{\lambda \lambda^{\prime}}(\tau) & \equiv & -i \Theta(\tau) 
\left\langle a_{\lambda}(\tau) a^{\dagger}_{\lambda^{\prime}}(0)
  \right\rangle_{\textrm{vac}}. 
\end{eqnarray}
\end{subequations}
To compute the Green functions, we differentiate Eq.\ 
(\ref{eq:green1}) with respect to $\tau$ and use the equations of
motion (\ref{eq:aieqm}) of the cavity operators $a_\lambda$.  This
yields the equations of motion of the Green functions
\begin{eqnarray}
\label{eq:geqm}
&&{\dot G}_{\lambda \lambda^{\prime}} (\tau) =  
\delta(\tau) G_{\lambda \lambda^\prime} (0) -i\omega_{\lambda} 
G_{\lambda \lambda^{\prime}} (\tau) \nonumber \\
&&\!\!\!\!\!- \int\limits_{0}^{\infty} \! 
d\omega^{\prime} \int\limits_{0}^{\tau} 
\! dt^{\prime} \; e^{-i\omega^{\prime}(\tau-t^{\prime})} \left[
 \mathcal{W}(\omega^{\prime})\mathcal{W}^{\dagger}(\omega^{\prime}) 
G (t^{\prime}) \right]_{\lambda \lambda^{\prime}}  
. \quad
\end{eqnarray}
There is no contribution from the noise term in Eq.\
(\ref{eq:aieqm}) as $\langle b_m(\omega,0) a^{\dagger}_{\lambda}
(0) \rangle_{\textrm{vac}} = 0$. The initial condition at
$\tau=0$ is $G_{\lambda \lambda^{\prime}}(0) = -i \delta_{\lambda
\lambda^{\prime}}$. Equation (\ref{eq:geqm}) is readily solved by
Fourier transformation. The result is
\begin{equation}
\label{eq:gfre}
G (\omega) = D^{-1}(\omega),
\end{equation}
where the non--Hermitean matrix $D$ was defined in Eq.~(\ref{eq:d}).
Substituting the result into the Fourier transform of Eq.\ 
(\ref{eq:2tcorr}), we obtain the field correlation function in the
frequency domain,
\begin{equation}
\label{eq:ct}
C_{ij}(\omega) = \frac{i\hbar \omega_{0}}{2} \sum_{\lambda
\lambda^{\prime}} u_{\lambda i}({\bf r}_0) u^{*}_{\lambda^\prime j}({\bf
r}_0) \left[D^{-1}(\omega) \right]_{\lambda \lambda^{\prime}},
\end{equation}
where we again made use of the rotating wave approximation to
replace $\sqrt{\omega_{\lambda} \omega_{\lambda^\prime}} \simeq
\omega_0$. The decay rate follows upon substitution of Eq.\ 
(\ref{eq:ct}) into Eq.\ (\ref{eq:gb}),
\begin{equation}
\label{eq:gct}
\gamma = -\frac{\omega_0}{\hbar} {\textrm{Im}} \! \bigg[ \sum_{ij} d_i
  d^{*}_j \sum_{\lambda \lambda^{\prime}} u_{\lambda i}({\bf
    r}_0) \! \left[D^{-1}(\omega_0)\right]_{\lambda \lambda^{\prime}} 
   u^{*}_{\lambda^\prime j}({\bf r}_0) \bigg] .
\end{equation}
The sum over modes may be simplified in the eigenbasis of the
non--Hermitean matrix $D$. In this basis, the double sum over the
mode functions ${\bf u}_\lambda$ reduces to a summation over the
left and right eigenmodes of the wave equation of the open cavity,
\begin{subequations}
\label{eq:gfin}
\begin{eqnarray}
\label{eq:gfina}
\gamma & = & \frac{\pi \omega_0 d^2}{\hbar}  
\rho ({\bf r}_0,\omega_0), \\
\rho ({\bf r}_0,\omega_0) &=& \frac{1}{\pi} {\textrm{Im}} 
\left[\sum_k \frac{ L_k^* ({\bf r}_0,\omega_0)  
R_k ({\bf r}_0,\omega_0) }{\omega_k - 
\omega_0 - i \frac{\Gamma_k}{2}} \right]. \quad \quad
\end{eqnarray}
\end{subequations}
Here, $\rho ({\bf r}_0,\omega_0)$ is the local density of modes at
the position of the atom, $L_k$, $R_k$ denote the component along
${\bf d}$ in the left and right mode $k$, and $\omega_k$ and
$\Gamma_k$ are the mode frequency and the mode broadening.

Equations (\ref{eq:gfin}) are the final result for the decay rate.
They describe spontaneous emission not only in cavities with
quasi--discrete modes but also in unstable resonators with strongly
overlapping modes. In the latter case, the left eigenfunctions of the
cavity may differ strongly from the corresponding right
eigenfunctions. Our result agrees with the decay rate derived in
Refs.\ \cite{Lam02,Bro01} using a field expansion in terms of
non--orthogonal modes.  Equations (\ref{eq:gfin}) have recently
\cite{Mis97} been used to calculate the distribution $P(\gamma)$ of
decay rates for a two--level atom inside a chaotic cavity. The local
density of modes also determines the photodissociation rate of small
molecules with chaotic internal dynamics \cite{Fyo98}. Our derivation
of the decay rate was based on the rotating wave approximation for the
system--and--bath Hamiltonian.  This approximation is valid for the
(typical) case in which the broadening of the resonator modes is much
smaller than the atomic transition frequency. When the mode broadening
is of the order of the transition frequency, one can still compute the
decay rate provided the coupling between the field and the atom is
sufficiently small. The calculation is done in Appendix \ref{app:C}.

\section{Conclusion}
\label{sec:conc}
We have presented an approach to the field quantization in
optical resonators. Our quantization scheme applies to fields of
arbitrary polarization and holds in the presence of an arbitrary
number of escape channels.

An attractive feature of our approach is that it is based on a field
expansion in terms of a set of {\em orthogonal} resonator and
channel modes. The creation and annihilation operators associated
with these modes obey canonical commutation relations. This is in
contrast to the non--standard commutation relations found in
alternative procedures that directly quantize the {\em
  non--orthogonal} modes of a non--Hermitean eigenvalue problem.
Non--Hermitean operators enter our approach only through the mode
{\em dynamics}.

In the case of weak damping our approach reduces to the well--known
field quantization of standard laser theory. Then the resonator
modes can be approximated by eigenmodes of a closed resonator. Each
mode is damped due to escape from the cavity. Deviations from this
simple dynamics show up when the damping rates are comparable with
the frequency spacing of the resonator modes. Then the Langevin
dynamics of the internal modes is coupled by damping and noise
inflicted by the external radiation field. Our field quantization
provides a unified description of both the regime of weak damping
and the regime of strong damping and spectrally overlapping modes.

In disordered dielectrics light scatters chaotically due to spatial
fluctuations of the dielectric constant. Chaotic scattering can be
included in our approach by assuming that the cavity dynamics is
described by a random matrix. The matrix is non--Hermitean since the
cavity is coupled to the external radiation field.  Averages over an
appropriate ensemble of random matrix theory yield statistical
information about the physical observables. In the present paper we
demonstrated the connection with random matrix theory for linear
optical media. Future application of the quantization technique to
random lasers may allow to extend the connection with random matrix
theory to non--linear optical media.

\begin{acknowledgments}
  We are indebted to H.--J.\ Sommers for important discussions
  concerning the Feshbach projector technique. We also enjoyed
  helpful discussions with F.\ Haake and D.\ V.\ Savin. This work
  has been supported by the SFB 237 ``Unordnung und gro\ss e
  Fluktuationen'' der Deutschen Forschungsgemeinschaft.
\end{acknowledgments}
\appendix

\section{Commutation relations for cavity and channels operators}
\label{app:A}

In this appendix we compute the (equal--time) commutation relations
for the cavity and channels position and momentum operators. The
reality condition on ${\bf A}$ and ${\bm \Pi}$ implies the following
relations between the cavity operators and their adjoints
\begin{eqnarray}
\label{eq:qprealres}
Q_{\lambda} &=& \sum_{\lambda^\prime}
\mathcal{N}^{\dagger}_{\lambda \lambda^\prime}
Q^{\dagger}_{\lambda^\prime}\\
P^{\dagger}_{\lambda} &=& \sum_{\lambda^\prime}
\mathcal{N}^{\dagger}_{\lambda \lambda^\prime}
P_{\lambda^\prime}.
\end{eqnarray}
Likewise, the channel operators are connected with their adjoints
via the relations
\begin{eqnarray}
\label{eq:qprealout}
Q_m(\omega) &=&  \sum_{m^\prime} \int \! d\omega^\prime \;
\mathcal{N}_{m m^\prime}^{\dagger}(\omega,\omega^\prime)
Q_{m\prime}^{\dagger}(\omega^\prime), \\
P_m^{\dagger}(\omega) &=& \sum_{m^\prime} \int \! d\omega^\prime \;
\mathcal{N}_{m m^\prime}^{\dagger}(\omega,\omega^\prime)
P_{m^\prime}(\omega^\prime).
\end{eqnarray}
The matrix elements
\begin{eqnarray}
\label{eq:usplit}
\mathcal{N}_{\lambda \lambda^\prime} &=& \int \! d{\bf r} \;
 {\bm \mu}_{\lambda}({\bf r}) \cdot
{\bm \mu}_{\lambda^\prime}({\bf r}), \nonumber \\
\mathcal{N}_{m m^\prime}(\omega,\omega^\prime) &=& \int \! d{\bf r} \;
 {\bm \nu}_{m}(\omega,{\bf r}) \cdot
{\bm \nu}_{m^\prime}(\omega^\prime,{\bf r}).
\end{eqnarray}
are the expansion coefficients of the mode functions ${\bm
  \mu}_\lambda$ (${\bm \nu}_m(\omega)$) in terms of the complex
conjugate functions ${\bm \mu}^*_\lambda$ (${\bm \nu}^*_m(\omega)$).
One can easily show that the matrices $\mathcal{N}$ are unitary,
symmetric and that they only couple degenerate modes.

The equal--time commutation relations follow easily from the
commutation relations (\ref{eq:qpcomm}) for the operators of the
total system and the completeness of the modes $|{\bm \phi}_m
(\omega)\rangle$. As an example we show that $[Q_\lambda,
P_m(\omega)] = 0$. Using the definitions (\ref{eq:qop}),
(\ref{eq:pop}) of $Q_\lambda$ and $P_m(\omega)$, we obtain
\begin{eqnarray}
\label{eq:qinpout}
[Q_\lambda,P_m(\omega)] &=& \sum_{m^{\prime} m^{\prime\prime}}
\int \! d\omega^{\prime} \int \! d\omega^{\prime\prime} \;
\alpha_{m^{\prime} \lambda}(\omega^{\prime})
\beta^{*}_{m^{\prime\prime} m} (\omega^{\prime\prime},\omega) \nonumber \\
& & \times [q_{m^{\prime}}(\omega^{\prime}),p_{m^{\prime\prime}}
(\omega^{\prime\prime})] \nonumber \\
&=& i \hbar \sum_{m^\prime} \int \! d\omega^{\prime} \;
\alpha_{m^\prime \lambda}(\omega^{\prime})
\beta^{*}_{m^\prime m}(\omega^\prime,\omega),
\end{eqnarray}
where we used the commutation relation (\ref{eq:qpcomm}). According
to Eq.\ (\ref{eq:univsplit}), the coefficients $\alpha$ and $\beta$
can be written as
\begin{eqnarray}
\label{eq:defal}
\alpha_{m^\prime \lambda}(\omega^\prime) & = & \langle
{\bm \mu}_{\lambda} | {\bm \phi}_{m^\prime}(\omega^\prime) \rangle ,
\nonumber \\
\label{eq:defbe}
\beta_{m^\prime m}(\omega^\prime,\omega) & = & \langle
{\bm \nu}_{m}(\omega) | {\bm \phi}_{m^\prime}(\omega^{\prime}) \rangle.
\end{eqnarray}
Substitution into the right hand side of Eq.\ (\ref{eq:qinpout}) yields
\begin{eqnarray}
\label{eq:appa1}
& &\sum_{m^\prime} \int \! d\omega^{\prime} \; 
\alpha_{m^\prime \lambda}(\omega^{\prime})
\beta^{*}_{m^\prime m}(\omega^\prime,\omega) \nonumber \\*
& &= \sum_{m^\prime} \int \! d\omega^{\prime}
\; \langle {\bm \mu}_{\lambda} | {\bm \phi}_{m^\prime}
(\omega^{\prime}) \rangle \nonumber \langle {\bm \phi}_{m^\prime}
(\omega^{\prime}) | {\bm \nu}_{m}(\omega) \rangle \nonumber \\*
& &= 0.
\end{eqnarray}
The calculation of all remaining commutators reduces to Eq.\ 
(\ref{eq:appa1}) or to one of the expressions
\begin{eqnarray}
\label{eq:appa2}
\sum_m \! \int \! d\omega \; \alpha_{m \lambda} (\omega) \alpha_{m
\lambda^\prime}^* (\omega)  &=&  \delta_{\lambda \lambda^\prime},\\
\label{eq:appa3}
\sum_{m^{\prime\prime}} \! \int \! d\omega^{\prime\prime} \;
\beta_{m m^{\prime\prime}} (\omega, \omega^{\prime\prime})
\beta_{m^\prime m^{\prime\prime}}^* (\omega^\prime,
\omega^{\prime\prime})  &=&  \delta_{m m^\prime} 
\delta(\omega -\omega^\prime). \nonumber \\*
\end{eqnarray}
One finds that the cavity operators have the commutation relations
\begin{eqnarray}
\label{eq:comres}
\begin{array}{lclcl}
{[Q_{\lambda},Q_{\lambda^\prime}]} &=&
[Q_{\lambda},Q^{\dagger}_{\lambda^\prime}] &=& 0, \\
{[P_{\lambda},P_{\lambda^\prime}]} &=&
[P_{\lambda},P^{\dagger}_{\lambda^\prime}] &=& 0, \\
{[Q_{\lambda},P_{\lambda^\prime}]} &=& i\hbar\,
\delta_{\lambda \lambda^\prime},\\
{[Q_{\lambda},P^{\dagger}_{\lambda^\prime}]} &=& i\hbar
\mathcal{N}^{*}_{\lambda \lambda^\prime}. \\
\end{array}
\end{eqnarray}
The channel operators have the commutation relations
\begin{eqnarray}
\label{eq:comout}
\begin{array}{lclcl}
[Q_{m}(\omega),Q_{n}(\omega^\prime)] &=&
[Q_{m}(\omega),Q^{\dagger}_{n}(\omega^\prime)] &=& 0, \\
{[P_{m}(\omega),P_{n}(\omega^\prime)]} &=&
[P_{m}(\omega),P_{n}^{\dagger}(\omega^\prime)] &=& 0, \\
{[Q_{m}(\omega),P_{n}(\omega^\prime)]} &=& i\hbar\,\delta_{m
  n}\,\delta(\omega-\omega^\prime), \\
{[Q_{m}(\omega),P^{\dagger}_{n}(\omega^\prime)]} &=& i\hbar
\mathcal{N}^{*}_{n m}(\omega^\prime,\omega), \\
\end{array}
\end{eqnarray}
and the cavity operators commute with all channel operators. This
shows that for each subsystem the operators $Q$ and $P$ behave like
the basic operators of position an momentum, respectively.

\section{The Hamiltonian}
\label{app:B}
We show how the Hamiltonian (\ref{eq:aaham}) is derived from the
Hamiltonian (\ref{eq:hqp}) that involves operators associated with
the eigenmodes of the total system. We separately treat the two
contributions to the Hamiltonian (\ref{eq:hqp}) involving integrals
over momentum and position operators, respectively. We start with
the contribution
\begin{equation}
\label{eq:appb1}
T = \frac{1}{2} \sum_m \int \! d\omega \;
p_m^\dagger (\omega)p_m(\omega) .
\end{equation}
Substitution of the representation (\ref{eq:qpsplit}) for $p_m
(\omega)$ reduces the right hand side to three integrals which can
be done using Eqs.\ (\ref{eq:appa1})--(\ref{eq:appa3}). The result
has the form
\begin{equation}
\label{eq:appb2}
T = \frac{1}{2}\sum_{\lambda} P^\dagger_{\lambda} P_{\lambda} +
\frac{1}{2} \sum_m \int \! d\omega \; P^{\dagger}_m(\omega) P_m(\omega).
\end{equation}
The second contribution
\begin{eqnarray}
\label{eq:qcont}
V = \frac{1}{2}\sum_m \int \! d\omega \;
\omega^2 q^{\dagger}_m(\omega)q_m(\omega)
\end{eqnarray}
is more difficult to compute due to the presence of the term
$\omega^2$ in the integral over frequency. Substituting the
representation (\ref{eq:qpsplit}) for $q_m(\omega)$ into Eq.\
(\ref{eq:qcont}), we obtain
\begin{eqnarray}
& &V= \frac{1}{2} \sum_{\lambda \lambda^{\prime}}
Q^{\dagger}_{\lambda} V^{(1)}_{\lambda \lambda^\prime}
Q_{\lambda^\prime} \nonumber \\*
& & + \frac{1}{2} \sum_{m^\prime m^{\prime\prime}}
\int \!  d\omega^{\prime} \int \!  d\omega^{\prime\prime} \;
Q_{m^\prime}^{\dagger}(\omega^\prime) V^{(2)}_{m^\prime
m^{\prime\prime}} (\omega^\prime,\omega^{\prime\prime})
Q_{m^{\prime\prime}} (\omega^{\prime\prime}) \nonumber \\*
& & + \frac{1}{2} \sum_{m^\prime} \sum_{\lambda} \int \! d\omega^\prime
\left( Q^{\dagger}_{m^\prime}(\omega^\prime) V^{(3)}_{m^\prime
\lambda} (\omega^\prime ) Q_{\lambda} + \mbox{h.c.} \right) ,
\end{eqnarray}
where $V^{(1)}$, $V^{(2)}$, and $V^{(3)}$ are integrals over the
coefficients $\alpha$ and $\beta$,
\begin{eqnarray}
\label{eq:defv1}
V^{(1)}_{\lambda \lambda^\prime} & = & \sum_m \int \! d\omega \;
\omega^2 \alpha_{m \lambda}(\omega) \alpha^{*}_{m \lambda^\prime}
(\omega) , \\
\label{eq:defv2}
V^{(2)}_{m^\prime m^{\prime\prime}} & = &
\sum_m \int \! d\omega \; \omega^2 \beta_{m^\prime m}
(\omega^\prime,\omega) \beta^{*}_{m^{\prime\prime}
m}(\omega^{\prime\prime},\omega) , \qquad\\
\label{eq:defv3}
V^{(3)}_{m^\prime \lambda} (\omega^\prime ) & = &
\sum_m \int \! d\omega \;\omega^2 \beta_{m^\prime m}
(\omega^\prime,\omega) \alpha^{*}_{m \lambda}(\omega).
\end{eqnarray}
We compute these integrals using relations that follow from the
eigenmode equation (\ref{eq:heveqsep}). Projecting this equation
onto the cavity states $\langle {\bm \mu}_\lambda|$ and the channel
states $\langle {\bm \nu}_m (\omega)|$, we obtain the set of coupled
equations
\begin{eqnarray}
\label{eq:exp1}
&& \!\!\!\!\!\left[ \frac{\omega^2}{c^2} - 
\frac{\omega^2_{\lambda}}{c^2} \right]
\alpha_{m \lambda}(\omega) = \frac{2}{c^2} \sum_{m^\prime}\! \int \!
d\omega^\prime W_{\lambda n}(\omega^\prime) \beta_{n m}
(\omega^\prime,\omega),\nonumber  \\*
\\
\label{eq:exp2}
&&\!\!\!\!\!\left[ \frac{\omega^2}{c^2} - 
\frac{\omega^{\prime 2}}{c^2} \right]
\beta_{nm}(\omega^\prime,\omega) = \frac{2}{c^2} \sum_{\lambda}
W^{*}_{\lambda m}(\omega^{\prime}) 
\alpha_{\lambda m}(\omega).\nonumber \\*
\end{eqnarray}
We used the definition $W_{\lambda m}(\omega) = (c^2 /2)
\langle {\bm \mu}_\lambda| L |{\bm  \nu}_{m} (\omega)
\rangle $ and the definitions (\ref{eq:defal}) of $\alpha$ and
$\beta$. We multiply both sides of Eq.\ (\ref{eq:exp1}) by
$\alpha^{*}_{\lambda^\prime m}(\omega)$ and then sum over $m$ and
integrate over $\omega$. This yields
\begin{eqnarray}
\label{eq:defv4}
V^{(1)}_{\lambda \lambda^\prime} &=& \omega^2_{\lambda} \sum_m
\int \! d\omega \; \alpha^{*}_{\lambda^\prime m}(\omega)
\alpha_{\lambda m}(\omega) \nonumber \\
& &+ 2 \sum_{m^\prime}
\int \! d\omega^\prime \; W_{\lambda m^\prime}(\omega^\prime) 
\nonumber \\*
& & \quad \quad \times\left[ \sum_m \int \! d\omega \; 
\alpha^{*}_{\lambda^\prime
    m}(\omega)\beta_{m^\prime m}(\omega^\prime,\omega) \right]. \quad \quad
\end{eqnarray}
The term in the square brackets vanishes according to Eq.\
(\ref{eq:appa1}). The remaining term on the right hand side can
be simplified using Eq.\ (\ref{eq:defal}), and Eq.\
(\ref{eq:defv4}) reduces to
\begin{eqnarray}
V^{(1)}_{\lambda \lambda^\prime} &=& \omega^2_{\lambda} \sum_m \int
\! d\omega \; \langle {\bm \mu}_{\lambda}|{\bm \phi}_m(\omega)\rangle
\langle {\bm \phi}_{m}(\omega)|{\bm \mu}_{\lambda^\prime}\rangle \nonumber \\*
&=& \omega^2_{\lambda} \, \delta_{\lambda \lambda^\prime}.
\end{eqnarray}
The expressions $V^{(2)}$ and $V^{(3)}$ can be computed in a
similar fashion. Combining results the second contribution to the
Hamiltonian takes the form
\begin{eqnarray}
\label{eq:qcont2}
V & = &  \frac{1}{2} \sum_{\lambda} \omega^2_{\lambda}
 Q^{\dagger}_{\lambda}Q_{\lambda} + \frac{1}{2}
\sum_m \int \! d\omega \; \omega^2
Q^{\dagger}_m(\omega) Q_m(\omega) \nonumber \\
&& + \sum_m \! \sum_{\lambda} \int \!
d\omega \left[  W_{\lambda m}(\omega) Q^{\dagger}_{\lambda} Q_m(\omega)
+ \mbox{ h.c.} \right].\qquad \quad
\end{eqnarray}
The sum of the contributions (\ref{eq:appb2}) and
(\ref{eq:qcont2}) yields the Hamiltonian (\ref{eq:aaham}).

\section{Spontaneous decay}
\label{app:C}
We show here how to derive the atomic decay rate $\gamma$ without
using the rotating wave approximation for the field
Hamiltonian. Our starting point is Eq.\ (\ref{eq:clap}), which
written in terms of the exact modes of the total system takes the
form \cite{Gla91}
\begin{eqnarray}
C_{ij}(t-t^{\prime}) &=& \Theta(t-t^\prime) \sum_m \int \! d\omega \; 
\frac{\hbar \omega}{2} \nonumber \\
&& \times f_{mi}(\omega,{\bf r}_0) f^{*}_{mj}(\omega,{\bf
  r}_0) \, e^{-i\omega(t-t^{\prime})}.
\end{eqnarray}
The Fourier transform of this equation is readily evaluated,
\begin{equation}
\label{eq:ctmu}
C_{ij}(\omega_0 + i\epsilon) = i \sum_m \int \! d\omega \frac{\hbar
  \omega}{2} \frac{ f_{mi}(\omega,{\bf r}_0) f^{*}_{mj}(\omega,{\bf
  r}_0)}{(\omega_0 - \omega) + i\epsilon}.
\end{equation}
Substitution into Eq.\ (\ref{eq:gb}) yields the golden rule result
\begin{eqnarray}
\label{eq:gexa}
\gamma &=& \frac{\pi}{\hbar^2} \sum_{ij} \sum_m \int \! d\omega \;
\hbar \omega \nonumber \\* & & \times \left[ d_{i} d^{*}_{j}
f_{mi}(\omega,{\bf r}_0) f^{*}_{mj}(\omega,{\bf r}_0) \delta(\omega -
\omega_0)\right].
\end{eqnarray}
The atom is located inside the cavity. Therefore, we can use the
mode expansion (\ref{eq:fsep}) to replace the modes of the total
system by the cavity modes,
\begin{eqnarray}
\label{eq:gexb}
& &\hspace*{-1.5pt}\gamma = \frac{\pi}{\hbar^2} \sum_{ij} \int \!
 d\omega \; \hbar \omega\, d_{i} d^{*}_{j} \nonumber \\
& &\hspace*{-1.5pt}\times \left[ \sum_{\lambda \lambda^{\prime}}
\sum_m u_{\lambda i}({\bf r}_0) u^{*}_{\lambda^{\prime}
j}({\bf r}_0)  \alpha_{\lambda m}(\omega)
  \alpha^{*}_{\lambda^{\prime}m}(\omega)\right]
\delta(\omega - \omega_0). \nonumber \\*
\end{eqnarray}
The quantity in square brackets is proportional to the local
density of states $\rho ({\bf r}_0,\omega)$,
\begin{eqnarray}
\label{eq:rloca}
-\frac{i\pi c^2}{\omega} \sum_{m}\sum_{\lambda \lambda^{\prime}}
{\bf u}_{\lambda}({\bf r}_0) \alpha_{\lambda m}(\omega)
\alpha^{*}_{\lambda^{\prime} m}(\omega) 
{\bf u}_{\lambda^{\prime}}({\bf r}_0)&& \nonumber \\
= 2i {\textrm{Im}} \, \langle {\bf r}_0| G_{\mathcal{QQ}}
\left(\frac{\omega^2}{c^2}\right) |{\bf r}_0\rangle. &&
\end{eqnarray}
Therefore, the atomic decay rate has the same form as in Eq.\
(\ref{eq:gfina}) but with a modified local density of modes
\begin{eqnarray}
\rho ({\bf r}_0,\omega_0) &=& \frac{2\omega_0}{\pi c^2}
{\textrm{Im}}
\left[\sum_k \frac{ l^*_k({\bf r}_0,\omega_0) r_k({\bf
    r}_0,\omega_0)}{ \sigma_k(\omega_0) -
    \left(\frac{\omega_0}{c}\right)^2} \right], \quad
\end{eqnarray}
where $l_k$ and $r_k$ are the left and right eigenmodes and
$\sigma_k$ the eigenvalues of the non-Hermitian operator
$L_{\textrm{eff}} (\omega_0)$. Equation (\ref{eq:gfin}) is recovered
in the rotating wave approximation. Then the eigenmodes of
$L_{\textrm{eff}}$ are simply related to the eigenmodes of the
non--Hermitean matrix $D^{-1}$, $l_k\approx L_k$ and $r_k\approx
R_k$, and the eigenvalues of $L_{\textrm{eff}} (\omega_0)$ can be
approximated by
\begin{equation}
\sigma_k \approx \left(\frac{\omega_k}{c}\right)^2 -
\frac{i\omega_0}{c^2} \Gamma_k,
\end{equation}
where $\omega_k$ and $\Gamma_k$ are the mode frequency and the mode
broadening.


\begin{thebibliography}{10}

\bibitem{Cao99} H.\ Cao {\it et al.}, Phys.\ Rev.\ Lett.\ {\bf 82},
  2278 (1999); H.\ Cao {\it et al.}, Phys.\ Rev.\ Lett.\ {\bf 84},
  5584 (2000); H.\ Cao {\it et al.}, Phys.\ Rev.\ Lett.\ {\bf 86},
  4524 (2001).
  
\bibitem{Cao00} H.\ Cao, J.\ Y.\ Xu, S.\ H.\ Chang, and S.\ T.\ Ho,
  Phys.\ Rev. E {\bf 61}, 1985 (2000).
  
\bibitem{Lin02} Y.\ Ling, H.\ Cao, A.\ Burin, M.\ A.\ Ratner,
  X.\ Liu, and R.\ P.\ H.\ Chang, Phys.\ Rev.\ A {\bf 64}, 063808
  (2001).

\bibitem{Fro98} S.\ V.\ Frolov, Z.\ V.\ Vardeny, and K.\ Yoshino,
  Phys.\ Rev.\ B {\bf 57}, 9141 (1998).
  
\bibitem{Wie00} D.\ S.\ Wiersma, Nature (London) {\bf 406}, 132
  (2000).

\bibitem{She90} {\it Scattering and Localization of Classical Waves
    in Random Media}, edited by P.\ Sheng (World Scientific,
  Singapore, 1990).
  
\bibitem{She95} P.\ Sheng, {\it Introduction to Wave Scattering,
    Localization, and Mesoscopic Phenomena} (Academic Press, San
  Diego, 1995).

\bibitem{Alb85} M.\ P.\ Van Albada and A.\ Lagendijk, Phys.\ Rev.\
  Lett.\ {\bf 55}, 2692 (1985); P.\ E.\ Wolf and G.\ Maret, {\it
    ibid.} {\bf 55}, 2696 (1985).

\bibitem{Imr97} Y.\ Imry, {\it Introduction to Mesoscopic Physics}
  (Oxford University Press, New York, 1997).
  
\bibitem{Gen91} A.\ Z.\ Genack and N.\ Garcia, Phys.\ Rev.\ Lett.\ 
  {\bf 66}, 2064 (1991).
  
\bibitem{Wie97} D.\ S.\ Wiersma, P.\ Bartolini, A.\ Lagendijk, and
  R.\ Righini, Nature (London) {\bf 390}, 671 (1997).  

\bibitem{And58} P.\ W.\ Anderson, Phys.\ Rev.\ {\bf 109}, 1492
  (1958); Philos.\ Mag.\ B {\bf 52}, 505 (1985).
  
\bibitem{Efe97} K.\ B.\ Efetov, {\it Supersymmetry in Disorder and Chaos} (Cambridge University Press, New York, 1997).
  
\bibitem{Ela98} B.\ Elattari, V.\ Kagalovsky, and H.\ A.\ 
  Weidenm\"uller, Phys.\ Rev. E {\bf 57}, 2733 (1998).

\bibitem{Hak84} H.\ Haken, {\it Laser theory}, 2nd edition (Springer,
  Berlin, 1984).
 
\bibitem{Sen59} I.\ R.\ Senitzy, Phys.\ Rev.\ {\bf 115}, 227 (1959);
Phys.\ Rev.\ {\bf 119}, 670 (1960).

\bibitem{Gar00} C.\ W.\ Gardiner and P.\ Zoller, {\it Quantum
    Noise}, 2nd edition (Springer, Berlin, 2000). 
  
\bibitem{Bar88} S.\ M.\ Barnett and P.\ M.\ Radmore, Opt.\ Commun.\ 
  {\bf 68}, 364 (1988).

\bibitem{Dut00a} S.\ M.\ Dutra and G.\ Nienhuis, Acta Physica
  Slovaca {\bf 50}, 275 (2000); J.\ Opt.\ B {\bf 2}, 584 (2000).

\bibitem{Lan73} R.\ Lang, M.\ O.\ Scully, and W.\ E.\ Lamb, Phys.\
Rev.\ A {\bf 7}, 1788 (1973).

\bibitem{Gla91} R.~J. Glauber and M. Lewenstein, Phys.\ Rev.\ A
{\bf 43}, 467 (1991).

\bibitem{Kno87} L.\ Kn\"oll, W.\ Vogel, and D.\ G.\ Welsch, Phys.\
Rev.\ A {\bf 36}, 3803 (1987); Phys.\ Rev. A {\bf 43}, 543 (1991).

\bibitem{Dal99} B.\ J.\ Dalton, S.\ M.\ Barnett, and P.\ L.\ Knight
J.\ Mod.\ Opt.\ {\bf 46}, 1315 (1999); S.\ Brown and B.\ J.\ Dalton,
J.\ Mod.\ Opt.\ {\bf 48}, 597 (2001).

\bibitem{You98} E.\ S.\ C.\ Ching, P.\ T.\ Leung, A.\ Maassen van
  den Brink, W.\ M.\ Suen, S.\ S.\ Tong, and K.\ Young, Rev.\ Mod.\
  Phys.\ {\bf 70}, 1545 (1998).

\bibitem{Lam99} C.\ Lamprecht and H.\ Ritsch, Phys.\ Rev.\ Lett.\
  {\bf 82}, 3787 (1999).

\bibitem{Lam02} C.\ Lamprecht and H.\ Ritsch, Phys.\ Rev.\ A {\bf 65},
023803 (2002).

\bibitem{Dut00b} S.~M. Dutra and G.~Nienhuis, Phys.\ Rev.\ A {\bf
62}, 063805 (2000).

\bibitem{Bro01} S.~A. Brown and B.~J. Dalton, J.\ Mod.\ Opt.\ {\bf
49}, 1009 (2002).

\bibitem{Gru96} T.\ Gruner and D.--G.\ Welsch, Phys.\ Rev.\ A {\bf
    54}, 1661 (1996).
  
\bibitem{Art97} M.\ Artoni and R.\ Loudon, Phys.\ Rev.\ A {\bf 55},
  1347 (1997).

\bibitem{Bee98} C.\ W.\ J.\ Beenakker, Phys.\ Rev.\ Lett.\ {\bf 81},
1829 (1998).

\bibitem{Mis00} E.\ G.\ Mishchenko, M.\ Patra, and C.\ W.\ J.\
Beenakker, Eur.\ Phys.\ J.\ D.\ {\bf 13}, 289 (2001).

\bibitem{Hac01} G.\ Hackenbroich, C.\ Viviescas, and F.\ Haake,
Phys.\ Rev.\ Lett.\ {\bf 89}, 083902 (2002).

\bibitem{foot1} We use the convention $\mathcal{M}^\dagger_{mn} (\omega,
  \omega^\prime)=[\mathcal{M}_{nm}(\omega^\prime ,\omega)]^*$.

\bibitem{Fes62} H.\ Feshbach, Ann.\ Phys.\ (N.Y.) {\bf 19}, 287 (1962).

\bibitem{Jac75} J.\ D.\ Jackson, {\it Classical Electrodynamics},
2nd edition (Wiley, New York, 1975).

\bibitem{Fan61} U.\ Fano, Phys.\ Rev.\ {\bf 124}, 1866 (1961).

\bibitem{Dit00} F.~M.\ Dittes, Phys.\ Rep.\ {\bf 339}, 215
  (2000).

\bibitem{Som02} D.\ V.\ Savin, V.\ V.\ Sokolov, and H.-J.\ Sommers,
cond--mat/0206176.
      
\bibitem{Coh92} C.\ Cohen-Tannoudji, J.\ Dupont-Roc, and G.\ 
  Gryberg, {\it Atom--Photon Interactions} (Wiley, New York, 1992).

\bibitem{Haa85} F.\ Haake and R.\ Reibold, Phys.\ Rev.\ A {\bf 32},
  2462 (1985).

\bibitem{Wal94} D.\ F.\ Walls and G.\ J.\ Milburn, {\it Quantum
Optics} (Sprin\-ger, Berlin, 1994).

\bibitem{Mah69} C.\ Mahaux and H.\ A.\ Weidenm\"uller, {\it
Shell--Model Approach to Nuclear Reactions} (North--Holland,
Amsterdam, 1969).

\bibitem{Bee97} C.\ W.\ J.\ Beenakker, Rev.\ Mod.\ Phys.\ {\bf
      69}, 731 (1997).

\bibitem{Fyo97} This assumption implies that $\omega_0$ is far from
  the threshold frequencies for the opening of further channels, see
  e.g.\ Y.\ V.\ Fyodorov and H.--J.\ Sommers, J.\ Math.\ Phys.\ {\bf
    38}, 1918 (1997).

\bibitem{Bar99} Similar Langevin equations have previously been derived
by P.\ J.\ Bardroff and S.\ Stenholm, Phys.\ Rev.\ A {\bf 60}, 2529
(1999); {\bf 61}, 023806 (2000).
  
\bibitem{Bibel} F.\ Haake, {\it Quantum Signatures of Chaos}, 2nd
  edition (Springer, Berlin, 2000).

\bibitem{Sok89} V.\ V.\ Sokolov and V.\ G.\ Zelevinsky, Nucl.\ Phys.\
A {\bf 504}, 562 (1989).

\bibitem{Cha97} J.\ T.\ Chalker and B.\ Mehlig, Phys.\ Rev.\ Lett.\
{\bf 81}, 3367 (1998).

\bibitem{Mis97} T.\ Sh.\ Misirpashaev, P.\ W.\ Brouwer, and C.\ W.\ 
  J.\ Beenakker, Phys.\ Rev.\ Lett.\ {\bf 79}, 1841 (1997).

\bibitem{Fyo98} Y.\ V.\ Fyodorov and Y.\ Alhassid, Phys.\ Rev.\ A {\bf
58}, R3375 (1998).

\end{thebibliography}
\end{document}